\documentclass[fleqn,usenatbib]{mnras}

\usepackage{newtxtext,newtxmath}

\usepackage[T1]{fontenc}

\DeclareRobustCommand{\VAN}[3]{#2}
\let\VANthebibliography\thebibliography
\def\thebibliography{\DeclareRobustCommand{\VAN}[3]{##3}\VANthebibliography}

\usepackage{graphicx}	
\usepackage{amsmath}
\usepackage{bm} 
\usepackage{comment}
\usepackage{multicol}   
\usepackage{subcaption}
\usepackage{hyperref}
\usepackage[dvipsnames]{xcolor}
\usepackage{xspace}

\newcommand{\tcc}{t_\mathrm{cc}}
\newcommand{\rc}{r_\mathrm{cl}}
\newcommand{\lism}{L_\mathrm{ISM}}
\newcommand{\rcrit}{r_\mathrm{crit}}

\newcommand{\dd}{\mathrm{d}}

\newcommand{\pow}[1]{\ifmmode{}^{#1}\else ${}^{#1}$\fi}

\newcommand{\Lya}{\ifmmode{\mathrm{Ly}\alpha}\else Ly$\alpha$\xspace\fi}

\newcommand{\cm}{\,\ifmmode{\mathrm{cm}}\else cm\fi}

\newcommand{\ergps}{\,\mathrm{erg}\,\mathrm{s}\ifmmode{}^{-1}\else ${}^{-1}$\fi}
\newcommand{\Mpch}{\,\mathrm{Mpc}\,\ifmmode h^{-1}\else $h^{-1}$\fi}
\newcommand{\snru}{\,\ifmmode{\mathrm{Myr}^{-1}}\else Myr${}^{-1}$\fi}
\newcommand{\kms}{\,\ifmmode{\mathrm{km}\,\mathrm{s}^{-1}}\else km\,s${}^{-1}$\fi}

\newcommand{\cl}{\ifmmode{\mathrm{cl}}\else cl\fi}
\newcommand{\fc}{\ifmmode{f_{\mathrm{c}}}\else $f_{\mathrm{c}}$\fi}

\newcommand{\mg}[1]{\textcolor{Purple}{\small [\bf MG: #1]}}

\title[Launch of Multiphase Winds]{The Launching of Galactic Winds from a Multiphase ISM}

\author[Hidalgo-Pineda, Gronke \& Grete]{
Fernando Hidalgo-Pineda,$^{1}$\thanks{E-mail: fernando@mpa-garching.mpg.de}
Max Gronke$^{2,1}$
and Philipp Grete$^{3}$
\\
$^{1}$Max Planck Institute for Astrophysics, Garching D-85748, Germany \\
$^{2}$Astronomisches Rechen-Institut, Zentrum für Astronomie, Universität Heidelberg, Mönchhofstraße 12-14, 69120 Heidelberg, Germany\\
$^{3}$Hamburger Sternwarte, Universität Hamburg, Gojenbergsweg 112, 21029 Hamburg, Germany\\
}

\date{Draft from \today}

\pubyear{\the\year{}}

\begin{document}
\label{firstpage}
\pagerange{\pageref{firstpage}--\pageref{lastpage}}
\maketitle



\begin{abstract}
Galactic outflows are a key agent of galaxy evolution, yet their observed multiphase nature remains difficult to reconcile with theoretical models, which often fail to explain how cold gas survives interactions with hot, fast winds.  
Here we present high-resolution 3D hydrodynamic simulations of hot outflows interacting with a multiphase interstellar medium (ISM), parameterised by its cold-gas volume filling fraction $f_v$, depth $\lism$, and characteristic clump size $\rc$.  
We identify a universal survival criterion, $f_v \lism \gtrsim \rcrit$, that generalises the classical single-cloud condition ($\rc > \rcrit$) and correctly predicts cold-gas survival across a wide range of ISM configurations -- including scale-free -- down to $\rc/\rcrit \sim 10^{-2}$.  
Remarkably, the resulting cold phase rapidly loses memory of the initial ISM structure and converges toward a self-similar clump mass spectrum following Zipf’s law ($\mathrm{d}N/\mathrm{d}m \propto m^{-2}$), implying that turbulent mixing and radiative condensation universally shape the morphology of multiphase outflows.  
Surviving gas assembles into extended plumes or confined cold shells of size $\sim \chi r_{\mathrm{cl,min}}$, which grow as mass is accreted from the hot phase.  
The interaction of an initially laminar wind with a clumpy ISM naturally drives turbulence in both phases, which we characterise through first-order velocity structure functions that follow a Kolmogorov scaling with an injection scale set by $\lism$, and velocity dispersions reaching $\sigma\sim c_{\rm s,cold}$.  
Finally, the areal covering fraction of the cold gas approaches unity even for $f_v \sim 10^{-3}$, while the volume filling fraction remains low, naturally explaining the ``misty'' nature of observed outflows.  
Together, these results link small-scale cloud–wind interactions to galaxy-scale feedback, and we discuss their implications for interpreting observations and for modelling multiphase galactic winds in larger-scale simulations.
\end{abstract}

\begin{keywords}
hydrodynamics -- galaxies: evolution -- ISM: structure -- ISM: jets and outflows -- Galaxy: kinematics and dynamics
\end{keywords}



\section{Introduction}
\label{sec:intro}
Galactic winds are a hallmark of star-forming galaxies and a central mechanism in their evolution. Observations show that the velocities and extent of these outflows correlate strongly with star formation rates \citep{thompson2024,rubin2014}, highlighting their connection to stellar feedback processes. Their presence is ubiquitous across cosmic time, playing a crucial role in shaping the phase distribution of the interstellar medium (ISM), regulating star formation, and redistributing baryons across galactic, circumgalactic (CGM) and even intergalactic scales \citep[see reviews by][]{veilleux2005galactic,peroux2018,thompson2024}.

There is broad consensus on the importance of galactic winds in cosmic evolution. First, spectroscopic observations reveal that outflows are often metal-enriched compared to the surrounding gas, confirming their ISM origin and their role in enriching the CGM and intergalactic medium (IGM) with metals \citep{lopez2020,veilleux2022}. Second, multi-wavelength data consistently show that winds are both multiphase and multiscale, exhibiting structures across wide spatial extents, central to both small and large-scales of the baryon cycle \citep{rupke2018,lopez2025}. Finally, cosmological simulations consistently require energetic feedback in the form of outflows to reproduce the observed stellar mass function, metallicity gradients, and gas fractions in galaxies \citep[]{Vogelsberger2014, Nelson2019, somerville2015}. Despite this, the precise coupling between supernova (SN)-driven outflows and the ISM remains poorly understood. Several simulation efforts have tried to model outflows in realistic ISM environments. Stratified disk simulations such as \textsc{TIGRESS} \citep{tigress2023} and \textsc{SILCC} \citep{Walch2015, girichidis2016} reproduce the multiphase structure and chemistry of the ISM and use stellar feedback to launch cold gas flows. However, these flows mostly take the form of fountains that fall back onto the galaxy. Their design also makes it difficult to follow large-scale expanding winds \citep[e.g.][]{martizzi2016}. Global disk models with embedded supernovae, such as \citet{schneider2024cgols}, improve the treatment of wind geometry but still lack the resolution needed to resolve the small-scale dynamics of cold cloudlets interacting with the hot wind.

X-ray and millimetric observations reveal that outflows have a complex phase structure, with cold gas ($\sim 10^4$ K) embedded within a hot wind ($\sim 10^6$ K) \citep[]{heckman1990, fisher2024}. Capturing this multiphase composition and dynamics is challenging not only for advanced simulations but also for simplified models. Classically, these simplified approaches study in detail the interaction of a single cold clump of gas with stellar-driven feedback, treating it as a test case for multiphase outflow formation. In this framework, the acceleration timescale for a cloudlet to be entrained by the wind is given by $t_\mathrm{acc} \sim \chi \rc / v_\mathrm{wind}$,
where $\chi$ is the cloud-to-wind density contrast, $r_\mathrm{cl}$ the cloud radius, and $v_w$ the wind speed. In contrast, the cloud destruction timescale due to hydrodynamical instabilities is $t_\mathrm{dest} \sim \chi^{1/2} \rc / v_\mathrm{wind}$, so for  $\chi = T_\mathrm{hot} / T_\mathrm{cold}$, $t_\mathrm{dest} \ll t_\mathrm{acc}$ suggesting clouds are destroyed before they can be accelerated \citep{klein1994,zhang2015}. Instead, observations suggest that cold gas is dynamically coupled to outflows and a fundamental outcome of feedback. Not reproducing this nature comes at the cost of miscapturing the properties of galaxies and their environments.

Extensive theoretical work has closely examined the survival of these cold clumps embedded in fast, hot winds. Several studies have shown that radiative cooling can alter the outcome significantly \citep{cooper2009,scannapieco2015,armillotta2017survival,farber2018impact}. In particlular, \citet{gronke2018} showed that in the strong cooling regime, where the cooling time of cold clouds is shorter than their disruption timescale, overdense clouds can resist destruction by hydrodynamic instabilities \citep[also see][for further investigations of the relevant cooling time]{LiHopkins2020,Kanjilal2020,Sparre2020}. This introduces a critical radius $r_\mathrm{crit}$, above which radiative gains exceed destructive mixing losses \citep{gronke2020}. However, the ISM is not composed of isolated spherical clouds but instead follows a scale-free distribution shaped by turbulence \citep{Elmegreen1997, federrath2016, beattie2025, grete2025}. In such media, the single-cloud model cannot fully explain the persistence of cold gas: neighboring clouds can shield one another from shear instabilities, cloud drag may be enhanced by certain geometries, and mixing can be suppressed in clumpier environments. Additional physics such as magnetic fields and viscosity further complicate this picture \citep{mccourt2015magnetized,hidalgo2024better,Brueggen2023}. Cold gas survival is therefore unlikely to depend on a single critical scale, but rather on the combined influence of geometry and gas properties.

Previous work such as \citet{bandabarragan2021} investigate the driving of multiphase outflows through the interaction of a fast-travelling wind with fractal-like structures and thus bridge the `classical' cloud crushing studies with the stratified disk simulations discussed above. However, the connection to the analytical `survival criterion' described above remains largely unexplored. \citet{antipov2025} identify and study the cooling regime of single clouds in multicloud set-ups, but their analysis is limited to two simulation runs from \citet{bandabarragan2021}, without a broad survey of the parameter space.


In this work, we will address this point by studying the impact of a fast, hot wind on a more realistic, multiphase ISM. Using high-resolution 3D hydrodynamical simulations, we aim to resolve the relevant length scales (e.g., $r_{\rm crit}$), and thus aim to generalize the survival criterion $t_{\rm cool,mix}<t_{\rm cc}$, which is based on a spherical cloud geometry, to provide a condition testable also for larger scale simulations.
Our goal is to identify the physical conditions under which multiphase outflows emerge, determine how ISM geometry affects cold gas survival, and characterise the dynamical and thermal coupling of winds to the ISM.

This paper is organized as follows. In Section~\ref{sec:methods}, we describe the simulation setup and the modeling of the ISM. Section~\ref{sec:results} presents the main results from the wind-tunnel runs, outlining the conditions for the emergence of multiphase outflows. In Section~\ref{sec:discussion:universal_criterion}, we derive a general criterion for cold gas survival, followed by an analysis of outflow properties and potential observational signatures in Section~\ref{sec:fv_fa}. We conclude by summarising our findings and discussing key limitations in Section~\ref{sec:conclusions}.

\begin{figure*}
    \centering
    \includegraphics[width=\textwidth]{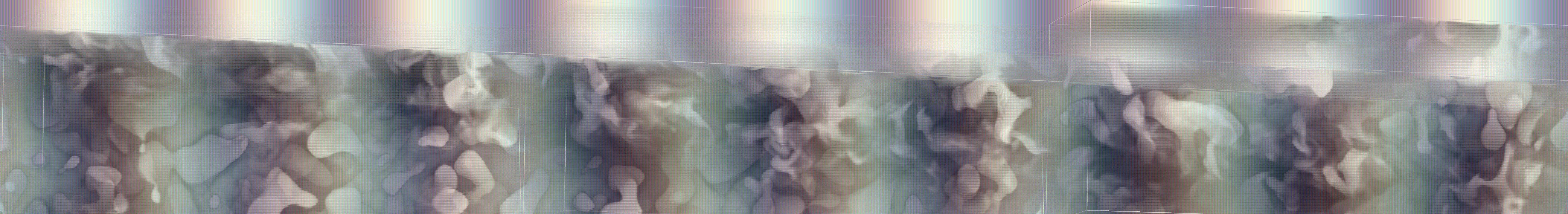}
    \caption{Volume rendering of our simulation set-up: a stellar wind with $\mathcal{M}_w = 1.5$ enters from the left, driving turbulent mixing as it moves through 1/4 of the cropped domain box. The rendering highlights regions of varying gas density. Bright yellow denotes gas that is overdense by a factor of 100 relative to the diffuse background phase (shown in light blue). Intermediate overdensities appear in muted colors, with black indicating regions where the overdensity is around 10. }
    \label{fig:Methods/long_box}
\end{figure*}

\section{Methods}
\label{sec:methods}

We conduct a suite of wind-tunnel simulations of hot, fast outflows interacting with idealised ISM density fields.  These simulations are designed to mimic a localised region of the galactic disk subjected to continuous wind-driven feedback. We extract a slab of multiphase ISM gas and expose it to a supersonic, high-temperature wind traveling parallel to the slab where we can study in detail the issue of entrainment. Figure~\ref{fig:Methods/long_box} shows a volume rendering of one quarter of a simulation box from our suite.

\subsection{ISM generation}
\label{methods:ism}

Both observations \citep{emelgreen2004,groves2023phangs}, simulations of the ISM \citep{Elmegreen1997,federrath2016,Naab2016} and simulations of turbulent media \citep{gronke2022survival,das2024magnetic, federrath2009fractal, beattie2025, grete2025} show cold matter arranges in clumpy filamentary structures, surrounded by a hot phase $10^6K$ gas. We generate this binary distribution of the ISM from two parameters, a volume filling fraction $f_v$ describing the abundance of cold gas relative to the tenuous phase, and a characteristic spatial lengthscale $\rc$ for the clumps. In principle, ISM turbulence is effectively scale-free at least over several orders of magnitude \citep{rathjen2025,federrath2009fractal}, Nevertheless, as a first step, we will study the behaviour for specific average clump sizes $\rc$ and later generalise it for a scale-free ISM setup.

For simplicity, we assume an isotropic distribution of gas, which we represent as a Poisson-like probability distribution such that each cell in our domain has a probability to be cold gas $\phi_\mathrm{(x,y,z)} \sim \mathcal{U}\{0,1\}$. We introduce spatial coherence by correlating neighboring cells to create an average clump size $\rc$, rather than treating gas cells as completely independent of their surroundings. Smoothing this Poissonian field with a Gaussian kernel (with standard deviation $\rc$)
\begin{equation}
    \tilde \phi_\textit{x, y, z} = \sum( G_{r_\mathrm{cl}} \ \ast \ \phi_\textit{x, y, z}),
\end{equation}
imprints the desired intrinsic scale.

For this field we can now employ a cutoff such that the final field will have a cold gas volume filling fraction $f_v$. Specifically, we can define a piece-wise function for the density field $\rho$ such that

\begin{equation}
    \displaystyle \rho(x,y,z) \equiv \begin{cases}
        \rho = \rho_\mathrm{hot} \textit{  if $\tilde{\phi}_\mathrm{x,y,z} < f_v$}\\
        \rho = \rho_\mathrm{cold} \textit{  if $\tilde{\phi}_\mathrm{x,y,z} \geq f_v$}
    \end{cases},
\end{equation}
or compactly written from a Heaviside function with offset $f_v$,
\begin{equation}
    \displaystyle \rho(x,y,z) = \rho_\mathrm{hot} \left [ 1 + (\chi - 1)\mathbf{H}(\tilde \phi - f_v ) \right ].
    \label{eq:ism_generation}
\end{equation}

\begin{figure}
    \includegraphics[width=\columnwidth]{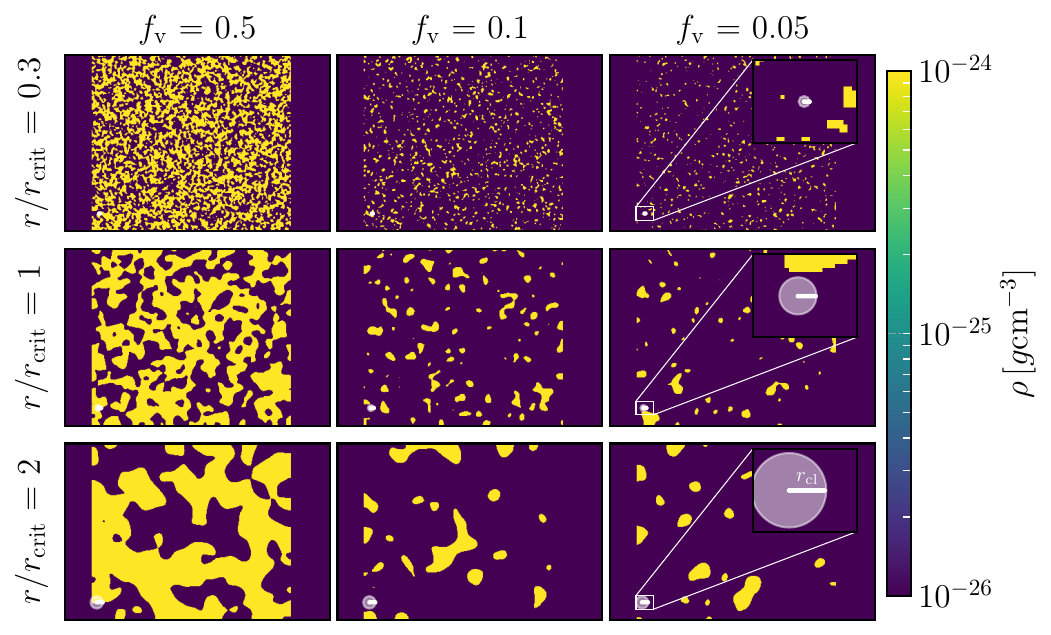}
    \caption{2D slices focused on our initial binary multiphase ISM fields where the wind enters from the left boundary (outside the shown region). We vary the volumetric cold gas filling fraction, $f_\mathrm{v}$, along the $x$ axis, and clump size $\rc$ along the $y$ axis, expressed as a fraction of the critical lengthscale for single  cloud survival, $r_\mathrm{crit}$. Colour-coded in yellow and navy the density values for cold and hot phases, respectively.}
    \label{Methods/fig:1}
\end{figure}

Figure~\ref{Methods/fig:1} represents the result of this process. The total cold mass of each set-up will also vary according the total depth of the ISM in the direction of the wind, $\lism$ (where $\lism$ is paralell to the $y$ axis). This parameter, along with $f_v$ and $\rc$, characterises our simulations. As we want to trace a universal criterion for the emergence of outflows, our suite of simulations span $\sim$ 4 orders of magnitude for all three variables (see details in Table~\ref{Methods/tab:simulations}). The resultant density field is used in our wind-tunnel simulation domain, where the $x$ and $z$ dimensions match the ISM extent, while the $y$ dimension is 2-3 times the ISM length ($\lism$).

\begin{table}
    \centering
    \caption{Parameters for our wind-tunnel ISM simulations: domain lengths parallel and perpendicular to the wind ($L_y$ and $L_z(L_x)$), cold gas volumetric fraction ($f_v$), clump size relative to the critical survival size ($\rc/r_\mathrm{crit}$), ISM length along the wind ($L_\mathrm{ISM}$), and  inflow Mach number ($\mathcal{M}_w$).
    }
    \begin{tabular}{lcccccr}
        \hline
        & $L_y[\rc]$ & $L_z (L_x)[\rc]$ & $f_v$ & $r_\mathrm{cl}/r_\mathrm{crit}$ & $L_\mathrm{ISM}[\rc]$ & $\mathcal{M}_w$ \\
        \hline
        1 & 96 & 32 & $10^{-1}$ & $10$ & 30 & 1.5\\
        2 & 96 & 32 & $10^{-1}$ & $10$ & 6 & 1.5\\
        3 & 96 & 32 & $10^{-2}$ & $10$ & 30 & 1.5\\
        4 & 96 & 32 & $10^{-3}$ & $10$ & 30 & 1.5\\
        5 & 96 & 32 & $10^{-1}$ & 1 & 30 & 1.5\\
        6 & 96 & 16 & $10^{-1}$ & 1 & 30 & 1.5\\
        7 & 112 & 48 & $10^{-1}$ & 1 & 20 & 1.5\\
        8 & 96 & 32 & $10^{-2}$ & 1 & 30  & 1.5\\
        9 & 864 & 32 & $10^{-2}$ & 1 &  300  & 1.5\\
        10 & 864 & 32 & $10^{-3}$ & 1 &  600  & 1.5\\
        11 & 4320 & 32 & $10^{-3}$ & 1 & 3000 & 1.5\\
        13 & 112 & 48 & $10^{-1}$ & 0.5 & 40 & 0.7\\
        14 & 176 & 48 & $10^{-2}$ & 0.5 & 40 & 0.7\\
        15 & 864 & 32 & $10^{-1}$ & 0.1 & 300 & 1.5\\
        16 & 96 & 32 & $10^{-1}$ & 0.1 & 30  & 1.5\\
        16 & 240 & 48 & $10^{-1}$ & 0.05 & 80 & 0.7\\
        17 & 432 & 32 & $10^{-1}$ & $10^{-2}$ & 300 & 1.5\\
        18 & 4320 & 32 & $10^{-1}$ & $10^{-2}$ & 3000 & 1.5\\
        \hline
    \end{tabular}

    \label{Methods/tab:simulations}
\end{table}

As a second setup, we generalise the ISM generation to produce a scale-free ISM (discussion in section~\ref{sec:overdense_examples}) ensuring the  mass distribution of clumps follows a mass profile density function $\dd N/\dd m \propto m^\mathrm{-2}$ found from both theory and observations of the ISM and turbulent structures \citep{gronke2022survival,Ilyasi_2025,tanfielding}.
From the definition of volume filling fraction, $f_v = N\rc^3/V_\mathrm{tot} \propto N$, and since $N \propto m^{-1}$ via simple integration of the mass distribution, we can establish the relation:

\begin{equation}
    f_v (r_\mathrm{cl})= f_\mathrm{v,max}\left( \frac{r_\mathrm{cl,min}}{r_\mathrm{cl}}\right)^3
    \label{eq:fv_relation}
\end{equation}
where $f_\mathrm{v,max}$ corresponds to our desired final volume filling fraction, majorly composed of the minimum cloud size $r_\mathrm{cl,min}$ for our ISM sample. Subsequent $\rc > r_\mathrm{cl,min}$ in the equation lead to lower individual $f_v$. Producing a series of realisations for different $\rc$ fields following this volumetric filling fraction relation recovers the clump mass power law. The final combined $f_v$ roughly corresponds to the sum of $f_v$ for each $\rc$ size. Essentially, the Heaviside function in \ref{eq:ism_generation} is replaced by
\begin{equation}
\mathbf{H}' = \sum_\mathrm{r = r_0}  \mathbf{H}\left(\tilde{\phi} \, - m/{r_\mathrm{cl}}^{3}\right)
\end{equation}
where $ m = f_\mathrm{v,max} \, r^3_\mathrm{cl,min}$. An example of the final mass distribution can be found in Appendix~\ref{app:ISM_generation}.

Note that, for all our runs, we satisfy the resolution criterion $\rc/d_\mathrm{cell} \geq 8$ which was found by previous `cloud crushing' studies to be sufficient in order to converge on the cold gas mass evolution \citep{gronke2020,Kanjilal2020}.

\subsection{Numerical implementation}
We perform wind-tunnel hydrodynamic simulations using the publicly available code \textsc{AthenaPK}\footnote{\textsc{AthenaPK} is an open-source, performance-portable code for astrophysical MHD simulations: \href{https://github.com/parthenon-hpc-lab/athenapk}{https://github.com/parthenon-hpc-lab/athenapk}.}, which implements finite-volume (magneto)hydrodynamic algorithms on top of the \textsc{Parthenon} framework~\citep{Parthenon}. We employ a formally second-order hydrodynamic finite-volume scheme with a predictor–corrector Van Leer integrator, Harten–Lax–van Leer with Contact (HLLC) Riemann solver, and piecewise parabolic reconstruction in primitive variables.  The performance portability of \textsc{AthenaPK} -- achieved through Kokkos \citep{kokkos} -- allows us to run on accelerator-based architectures and perform large-domain simulations on a static grid while strictly satisfying the aforementioned mass-growth resolution criterion of $\rc/d_\mathrm{cell} \ge 8$.

Our simulation meshes are structured as a rectangular domain with transverse lengths of $256$ or $128$ cells ($L_\perp /\rc = 32$ and 18, respectively) and a longitudinal dimension aligned with the wind direction ($y$-axis), typically extending to $L_y \sim 3 \lism$ (see Table~\ref{Methods/tab:simulations} for details). Both transverse lengths are sufficient to capture the dynamics, though we adopt the $256$ cell width as the fiducial resolution (see Appendix~\ref{app:resolution} for a resolution test). Our multiphase ISM density field realisations assign a temperature of $T = 10^6$ K  to the background (hot) phase, and a density of $\rho_\mathrm{hot} = 10^{-26}\text{g\,cm}^{-3}$, while the overdense (cold) phase is set to $T = 10^4$ K and $\rho_\mathrm{cold} = 10^{-24}\text{g\,cm}^{-3}$. Note it is only the relative overdensity of the gas at these two temperatures what affects the dynamical timescales, producing a self-similar solution regardless of the specific density values \citep[see \S~2.2 in][for an explanation on self-similarity]{Dutta2025}.

The ISM slab, of length $\lism$, is placed 8$\rc$ from the upstream $y$ boundary of the domain. From this boundary, we continuously inject a mildly transonic ($\mathcal{M} \approx 1.5$) or subsonic ($\mathcal{M} \approx 0.7$) hot wind in the positive $y$-direction. No initial turbulent velocities are imposed to either of the two phases, i.e.,  the hot wind is initially fully laminar. Periodic boundary conditions are applied in the transverse directions, $x$ and $z$, and a convergence study assessing potential artifacts from this setup is presented in Appendix~\ref{app:resolution}.

To compute radiative losses, we employ the \citet{townsend2009} radiative cooling algorithm, using the collisional ionization equilibrium (CIE) cooling tables from \citet{gnat_sternberg2007} for solar metallicity. To mimic heating from the UV background, we disable cooling above $6 \times 10^5$ K, and impose a temperature floor of $10^4$ K. To optimize computational efficiency and better track the motion of dense material, we perform simulations in a frame boosted fashion alike previous works \citep{mccourt2015magnetized,scannapieco2015,Dutta2019} by computing the the cold gas mass-weighted velocity
\begin{equation}
\langle v_\mathrm{cold} \rangle = \frac{ \int  \rho (T < 2T_\mathrm{cold})v_y dV}{\int  \rho (T < 2T_\mathrm{cold}) dV}
\end{equation}

with $\rho, T, V$ as density, temperature and volume, and $T_\mathrm{cold} = 10^4 K$, where our simulation set-up is on the frame-reference of cold gas, with $\langle v_\mathrm{cold} \rangle = 0$.

\section{Results}
\label{sec:results}

\begin{figure*}
    \centering
    \includegraphics[width=\textwidth]{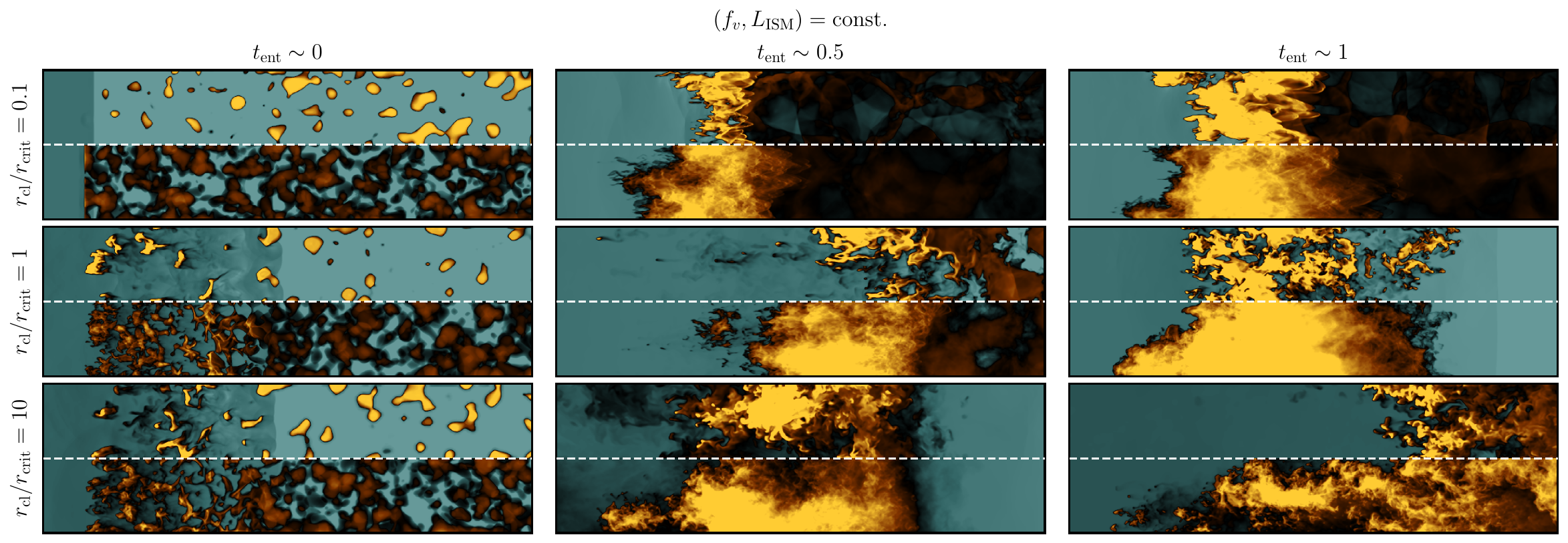}
  
\end{figure*}  

\begin{figure*}
    \centering
    \includegraphics[width=\textwidth]{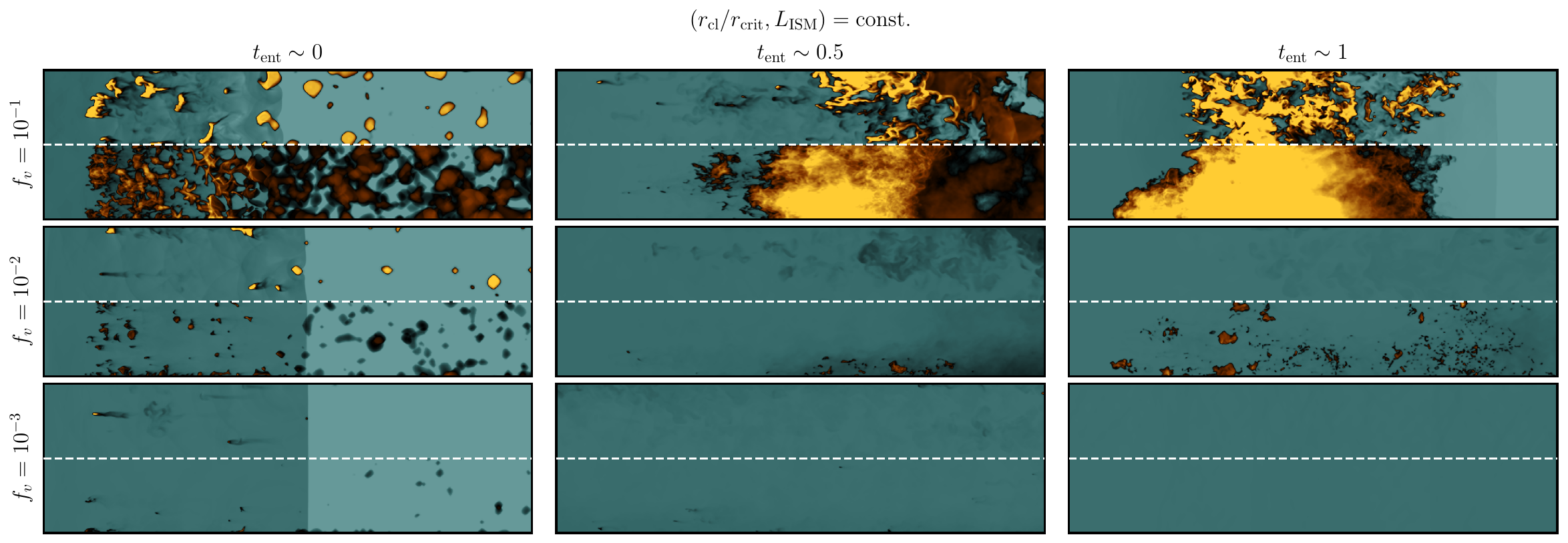}
  
\end{figure*}  
\begin{figure*}
    \centering
    \includegraphics[width=\textwidth]{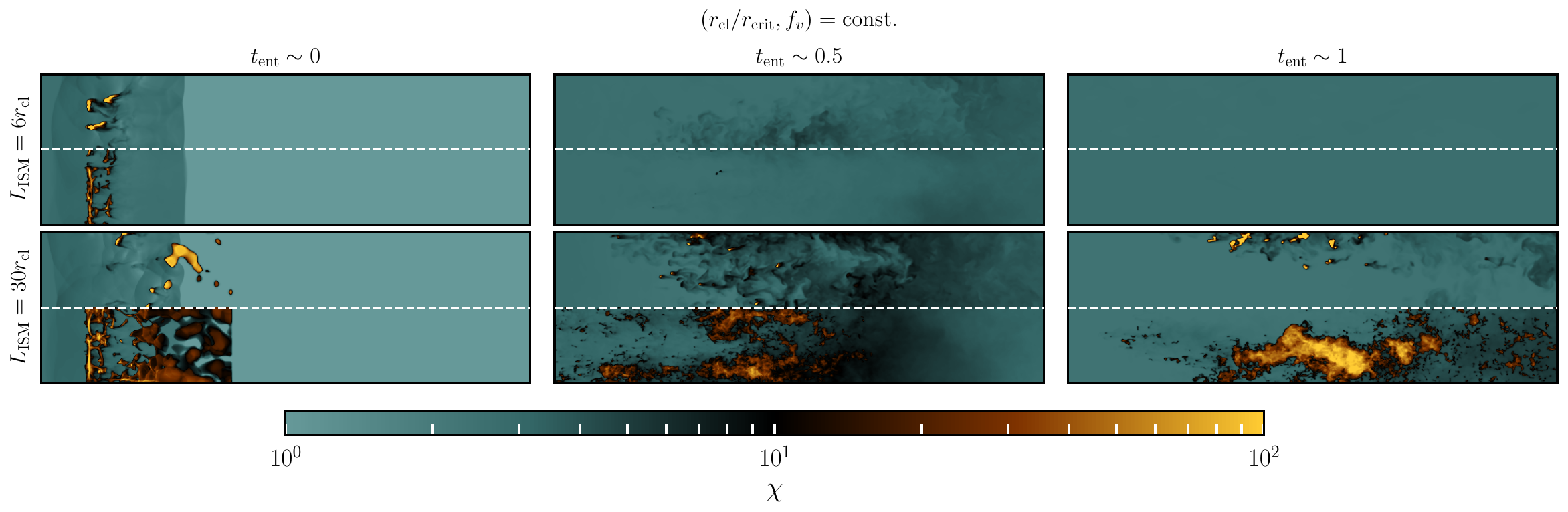}
    \caption{Density projection ($\int \rho(z) \mathrm{d}z / (\ell_z \rho_{\rm hot})$) at the time of wind-ISM interaction (left column), mid-way through entrainment (middle) and at time of entrainment
    (right) for different simulations. Each panel is divided into two halves by a dashed white line, where the top uses a total integration depth $\ell_z = \rc$, and bottom $\ell_z = L_\mathrm{z, box}$. {\bf Top (rows 1-3):} runs of varying $\rc$ and fixed $f_v=0.1, \lism = 300\rc$.
    {\bf Middle (rows 4-6):} runs of varying $f_v$ and fixed $\lism = 300\rc, \rc = \rcrit$.
    {\bf Bottom (rows 7-9):} 2 runs of varying $\lism$ and fixed $\rc=\rcrit,f_v = 0.1$.
    Note that we do not show the entire simulation domain but instead focus on the upstream boundary of the cold gas with the hot wind, and use a length of $L_y \approx 100 \rc$ along the $y$ axis.}
    \label{Results/fig:multiplot}

\end{figure*}

\subsection{Cold gas structure and morphology}
\label{sec:results/multiplot}
In Figure~\ref{Results/fig:multiplot} we show simulations where two of the three ISM control parameters ($f_v, \rc, \lism$) are fixed and the third varied. The characteristic destruction timescale for any of these systems corresponds to the canonical cloud-crushing timescale \citep{klein1994,scannapieco2015}
\begin{equation}
    t_\mathrm{cc} = \chi^{1/2} \frac{\rc }{v_\mathrm{wind}}.
\end{equation}

In Fig.~\ref{Results/fig:multiplot} as well as hereafter, we refer to the dimensionless timescale 
\begin{equation}
    \tau \equiv \frac{t - 0.5 \,t_\mathrm{sh}}{t_\mathrm{cc}}
\end{equation}
where $t_\mathrm{sh} = \lism/v_\mathrm{wind}$.
This is a convenient choice when comparing runs of different ISM depths, as the effects of the wind on cold gas mass and speed only become evident after having transversed half the total ISM depth. Similarly, the entrainment time can be derived from momentum conservation, $\rho_w A_\mathrm{cl}v_wt_\mathrm{ent} \approx A_\mathrm{cl}\left(\rho_\mathrm{cl} f_v \lism + \rho_w (1-f_v)\lism\right)$, with $A_\mathrm{cl}, v_w, \rho$ as area of the cloud, velocity of the wind and density (from either the hot $\rho_w$ or cold $\rho_\mathrm{cl}$ gas component). In contrast to single-cloud models, assuming $\rho_\mathrm{cl} \gg \rho_w$, the entrainment time of an ISM run
\begin{equation}
    t_\mathrm{ent} \approx \frac{\lism}{v_w} \left(\chi f_v + 1\right) = t_\mathrm{sh}(\chi f_v + 1),
\end{equation}

is directly proportional to the total cold gas mass depth and can be orders of magnitude larger than $\tcc$.

Figure~\ref{Results/fig:multiplot} shows that final structure (rightmost column) of the outflows varies significantly depending on the initial conditions. In the top subfigure (top 3 rows), $f_v$ and $\lism$ are fixed at 0.1, $300\rc$ resepectively, while $\rc/\rcrit$ is varied as 0.1, 1, and 10 from top to bottom. In this case, all three runs exhibit similar evolutionary behaviour. Initially, a clumpy multiphase distribution of the ISM is present as described in the initial conditions. At approximately half the entrainment time ($t_\mathrm{ent} \sim 0.5$), some cold clouds begins to coagulate, forming larger clumps while simultaneously losing some of their mass, represented by the black coloration, which indicates the presence of gas mixing. The top panel, for example, shows clumpier structures at this time with respect to the other panels, with substantial mixing taking place. By the entrainment time ($t_\mathrm{ent} \sim 1$), all runs exhibit a large fraction of cold mass. The bottom half section for each of these panels (below the dashed white line), showing the projected mass over the full box size, show the formation of regions where the local concentration of cold gas is higher, while at the same time, exhibiting a clumpy nature when integrated over shorter lengthscales (top half of the panels).

In the central subfigure of Fig.~\ref{Results/fig:multiplot} (middle 3 rows), we fix $\rc = \rcrit$ and $\lism = 300 \rc$ while varying the volume filling fraction $f_v$ from top to bottom: $f_{v}=10^{-1}$, $10^{-2}$, and $10^{-3}$.
At high filling fractions ($f_v=10^{-1}$, top row), the evolution resembles the runs shown in the subfigure above. Entrainment drives coagulation that competes with mixing, forming a coherent blob where cold gas preferentially survives.
At intermediate filling fractions ($f_v = 10^{-2}$, middle row), the evolution follows a similar pattern with continued mixing and coagulation. However, by the final snapshot, the surviving cold gas is mistier and has a smaller projected area than for $f_v = 0.1$. Comparing the rightmost column for both $f_v = 10^{-1}$ and $f_v = 10^{-2}$ reveals that lower filling fractions produce mistier structures, while higher filling fractions create more distinct, compact clumps. Such differences may help observers identify which galactic regions preferentially drive multiphase gas to galaxy outskirts based on whether the cold phase appears misty or clumpy (e.g., \citealt{chen2023}).
At low filling fractions ($f_v = 10^{-3}$, bottom row), the cold gas does not survive the interaction with the hot wind and no cold mass remains by the time of entrainment.

Figure~\ref{Results/fig:multiplot} also demonstrates that $\lism$ contributes to determine the phase of multiphase outflows. In the last subfigure (bottom 2 rows), we show two runs of $f_v = 0.1$ and $\rc = \rcrit$, but of $\lism = 6$ and $30\rc$, top and bottom, respectively. Although the shallower ISM does not retain any of its initial cold mass after the wind interaction, increasing it by a factor of 5 shows that it forms large cold phase structures.

\subsection{Evolution of gas phases}
\label{sec:dynamical_ev}

\begin{figure}
    \centering
    \includegraphics[width=\columnwidth]{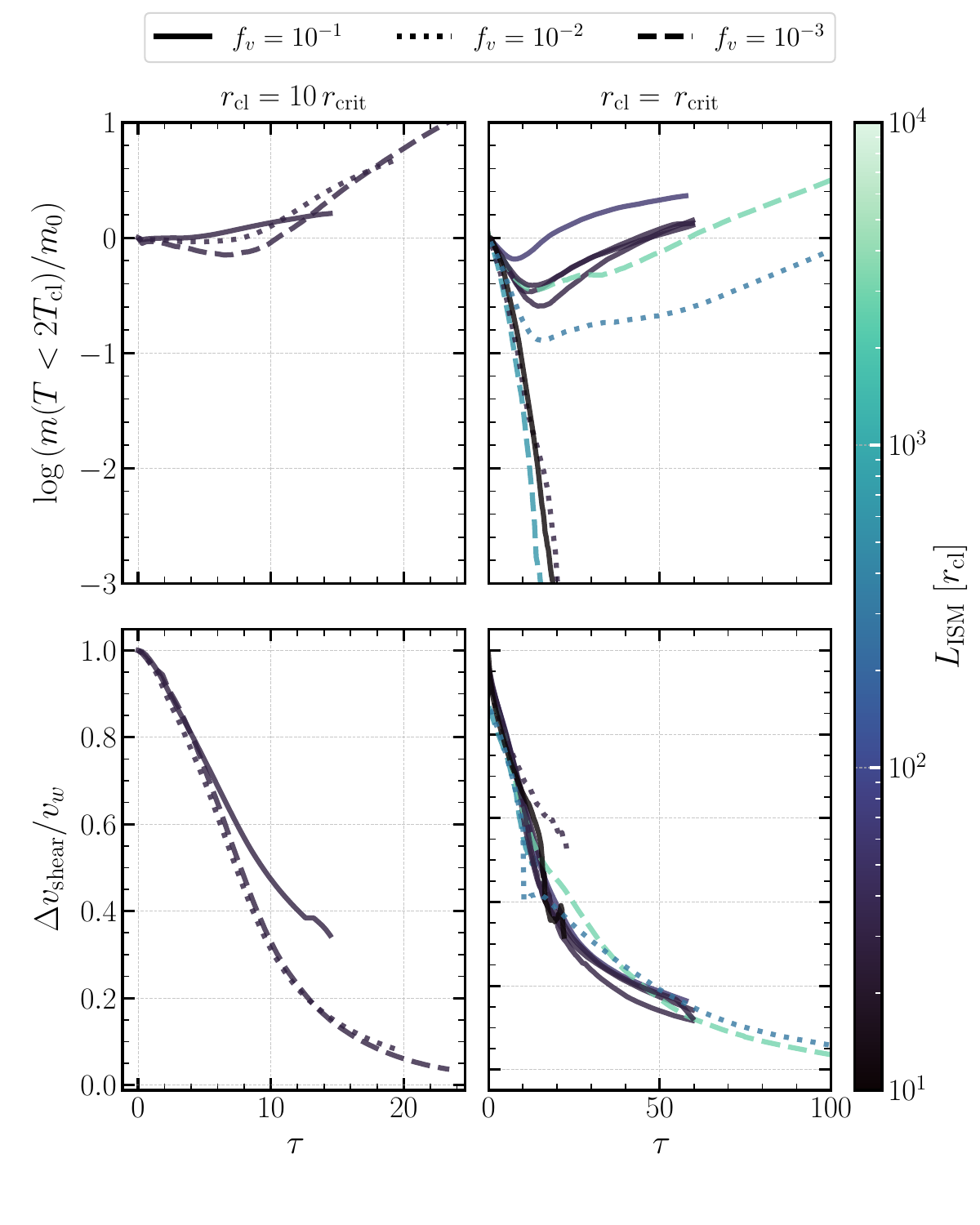}
    \caption{Total cold gas mass (top) and shear speed (bottom) evolution as function of their dimensionless temporal variable, $\tau$ (see \S~\ref{sec:results/multiplot}). Left panels display  runs for $\rc/r_\mathrm{crit} = 10$ (where the classical survival criteria of cold gas is strictly satisfied), and $r/r_\mathrm{crit} = 1$ for the right panels. Linestyles represent the volumetric filling fraction of cold gas in the ISM, and runs are colour-coded by the total depth of the initial density field in units of clump size $\rc$. }
    \label{Results/fig:mass_evol}
\end{figure}

\begin{figure*}
    \centering
    \includegraphics[width=\textwidth]{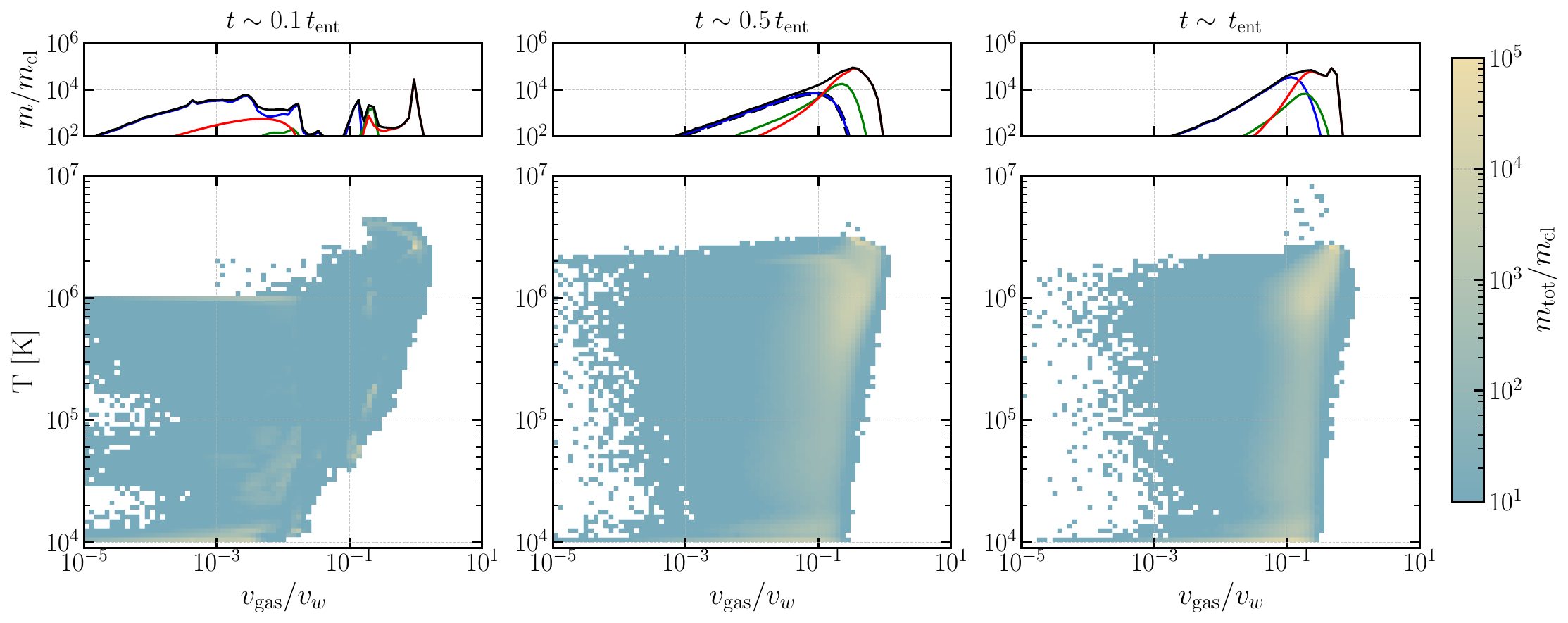}
    \caption{Top: mass carried per velocity bin during the wind-ISM interaction ($t \sim 0.1 \, t_\mathrm{ent}$, left), mid-way through entrainment ($t \sim 0.5 \, t_\mathrm{ent}$, middle) and at time of entrainment ($t \sim \,t_\mathrm{ent}$, right). In black, total mass velocity curve; blue, green and red show the mass curves for temperature cuts $T \leq 10^5, 10^5  \leq T \leq 10^{5.8}, T \geq 10^{5.8}$ K, respectively. We perform a least-squares fit (see S\ref{sec:phase_plot}) to the mass distribution of cold gas in the middle panel (dashed black). Bottom: phase diagrams for gas temperature (y-axis) and bulk flow speed along the wind direction (x-axis) as a fraction of wind speed $v_w$, weighted by phase bin mass (colourbar). This corresponds to run $\rc/r_\mathrm{crit} = 1, f_v = 0.1$ and $\lism = 30\rc $ .}
    \label{Results/fig:vel_phase}
\end{figure*}

\subsubsection{Cold mass evolution}

In order to place tighter constraints on the formation of multiphase outflows, we quantitatively analyse the evolution of cold gas mass over time. Figure~\ref{Results/fig:mass_evol} shows the time evolution of cold gas mass (top) and shear velocity relative to the wind (bottom) for two representative cases: one that strictly satisfies the classical survival criterion ($\rc/r_\mathrm{crit} = 10$) and one borderline case ($\rc/r_\mathrm{crit} = 1$). Linestyles denote volume filling fractions, with solid for $f_v = 10^{-1}$, dotted for $f_v = 10^{-2}$ and dashed for $f_v=10^{-3}$. The runs are colour-coded by ISM depth relative to the initial clump size (the full set of initial conditions is provided in Table~\ref{Methods/tab:simulations}).

The left panel of Fig.~\ref{Results/fig:mass_evol} shows that cold gas survives across all values of $f_v$ and $\lism$. The cold gas mass increases over time for all ISM depths and volume filling fractions. Besides, the bottom panel reveals that this mass growth is accompanied by acceleration to the wind speed. In other words, for the simulations shown in the left column, only the intrinsic initial clump size $\rc$ affects cold gas survival — neither $f_v$ nor $\lism$ plays a significant role. 
Notice that the entrainment time is $ t_\mathrm{ent} \sim t_\mathrm{sh} (\chi f_v + 1)\   $ as discussed in \S\ref{sec:results/multiplot}.

In the $\rc/r_\mathrm{crit} = 1$ case (right panels of Fig.~\ref{Results/fig:mass_evol}), survival is no longer guaranteed. Solid lines ($f_v = 0.1$) show mass ablation at early times. For instance, simulations with ISM depths of $10\rc$ (black lines) rapidly lose most of their mass within a few $\tau$. However, as we increase the ISM depth to $30\rc$ (dark purple) or more, mass growth resumes.
At lower $f_v$ (e.g., $10^{-2}$), a similar trend is observed: shallower depths ($\sim 10\rc$) lead to destruction, while deeper ISM columns of gas ($\sim 100\rc$) allow the cold phase to survive and eventually grow. For instance, the blue dotted line with $\lism = 300 \rc$ and $f_v = 10^{-2}$ shows mass recovery and growth after an initial drop to 10\% of the starting mass. Expectedly, the velocity evolution (bottom row panels) for the survival cases in this regime are similar among them and to the evolution of ISM runs with $\rc/r_\mathrm{crit} = 10$. 

This same behaviour holds for volumetric fractions as low as $f_v = 10^{-3}$ (dotted lines in Fig.~\ref{Results/fig:mass_evol}): while some cases (navy line) experience destruction, others (light blue curve) survive and grow. For the latter case, the ISM depth exceeds the individual clump size $\rc$ by three orders of magnitude.  

\subsubsection{Temperature-velocity evolution}
\label{sec:phase_plot}
Figure~\ref{Results/fig:vel_phase} shows the general temperature and velocity evolution of all gas cells at three different snapshots: initial shock-ISM interaction ($t_\mathrm{ent} \approx 0.1$), mid-entrainment ($t_\mathrm{ent} \approx 0.5$), and near full entrainment ($t_\mathrm{ent} \approx 1$), from left to right, respectively, for a simulation with $\left(f_v, \rc, \lism\right) = (10^{-1}, r_\mathrm{crit}, 30 r_\mathrm{cl})$.

At $t_\mathrm{ent} = 0.1$, the top panel shows that as the wind encounters the ISM, the majority of the cold phase remains slow in speed, while the inflowing hot phase is predominantly faster than the rest of the ISM gas. The bottom panel reveals that the wind has only begun to interact with interstellar gas, producing three main phase populations in dim yellow: the fast-travelling wind ($T > 10^6 , K$ and $v \simeq v_\mathrm{w}$), an equally tenuous static phase with $v \ll v_w$, and a static $10^4 \, K$ ISM component. Thermally-unstable gas at intermediate temperatures arises directly from early-disrupted ISM regions heated by the shock front, and is therefore scarce at this stage in evolution. The spread in velocities of this intermediate gas is a result of turbulent mixing in clouds swept by the wind, with $v_\mathrm{gas} \sim v_w$, and freshly disrupted clumps that mix into $\sim10^5$ K gas at $v \sim 0 $, therefore filling a wide velocity range below $v_w$.

At half the entrainment time ($t_\mathrm{ent} = 0.5$), the top panel shows that the cold phase velocity is now well described by a Schechter function $m(v)/m_\mathrm{cl} \approx A\, v/v_c \exp{(-(v/v_c)^2)}$ (with $A=10^4$ and $v_c = 0.1v_w$ shown by the thick dashed black line), with the bulk of its mass travelling at high speeds and a tail of accelerating gas extending down to $v/v_w \sim 10^{-3}$. During this time, however, the hot phase still carries most of the momentum, followed closely by mixed gas which contains more overall mass and travels faster than the cold phase of the outflow. The bottom panel shows that the initially unstable, low-velocity $10^5$ K gas now occupies a narrower $v_\mathrm{gas}$ distribution closer to $v_w$, with gas at these temperatures mainly produced through mixing from surviving clumps at higher speeds. Cold gas forms a distinct elongated region centred at $T \sim 10^4$ K and $v \sim 0.1v_\mathrm{w}$. The spread in $v_y$ for the cold phase is a distinctive feature of multicloud systems. Unlike single-cloud simulations, clouds in multicloud environments undergo differential entrainment: clouds at the leading edge are accelerated early by the wind, while those farther downstream have not yet been reached. Additionally, low-velocity fragments from ablated clouds can coagulate into larger clumps that retain these initially low velocities. These reformed clumps remain slow-moving until radiative cooling becomes efficient enough to enable their entrainment. This coagulation process, also illustrated in the central panel of Figure~\ref{Results/fig:multiplot}, is absent in single-cloud simulations.

At $t_\mathrm{ent} = 1$, the top panel shows that once both cold and hot phases are approximately co-moving, the mixed gas mass becomes negligible, and the cold phase carries most of the wind’s momentum. The bottom panel indicates that low-speed cold gas has mostly vanished, with negligible gas remaining at $v \lesssim 10^{-3}v_\mathrm{w}$. The surviving cold ($10^4$ K) and hot ($10^6$ K) components now co-travel roughly at the hot phase transonic speed $\mathcal{M}_\mathrm{w} \sim 1.5$. Gas cells at $10^5$ K sit below $0.01 , m_\mathrm{clump}$ and cluster around $v \sim v_\mathrm{w}$, consistent with turbulent radiative mixing layer theory in the fully entrained single-cloud picture. This mixed gas is however four orders of magnitude lower in mass than the two main wind phases.\\

Overall, while the gas structure in our set-ups is distinctly more intricate than idealised single-cloud models, the evolution of the system as a whole is broadly consistent with the expected dynamic and thermal evolution of individual clumps. Nevertheless, the formation of multiphase outflows does not strictly follow from the traditional single-cloud survival criterion. In the following section, we examine how this criterion must be modified to accurately predict multiphase gas formation.

\subsection{Cold gas survival}
\label{sec:results/survival_derivation}
We define a unique ISM from three parameters, $(f_v, \rc, \lism$; see section~\ref{methods:ism}). 
Analysing the mass evolution of our set of simulations showed that if $\rc \leq \rcrit$, estimating a threshold for the formation of a multiphase outflow requires a non-trivial combination of these three values, hinted by the alternate survival status for similar runs in the right panel of figure~\ref{Results/fig:mass_evol}. 

Analogous to the single cloud survival criterion $\rc > \rcrit$ we can compare the average cold gas length through the ISM, $f_v \lism$, to the critical cloud radius and conjecture that 
\begin{equation}
\alpha f_v \, \lism > \rcrit
\label{eq:surv_criterion}
\end{equation}
(where $\alpha$ is a fudge factor of order unity) leads to cold gas survival also in more complex ISM gas distributions.

Equation~\ref{eq:surv_criterion} gives a clear threshold for multiphase outflows, where the volumetric filling fraction $f_v$, and ISM depth $\lism$ both influence survival. Notice how the survival does not directly depend on the original clump size $\rc$ anymore. This is consistent with our results, as for all $\rc \leq \rcrit$, we always find a regime where the ISM survives the interaction (we comment on the limits of our criterion in section~\ref{sec:discussion:universal_criterion}). This equation simultaneously captures the single-cloud criterion for clouds $\rc > r_\mathrm{crit}$, as setting $f_v \approx 1$ and $\lism = \rc$ in equation~\ref{eq:surv_criterion} automatically satisfies the inequality.

\begin{figure}
    \centering
    \includegraphics[width=\columnwidth]{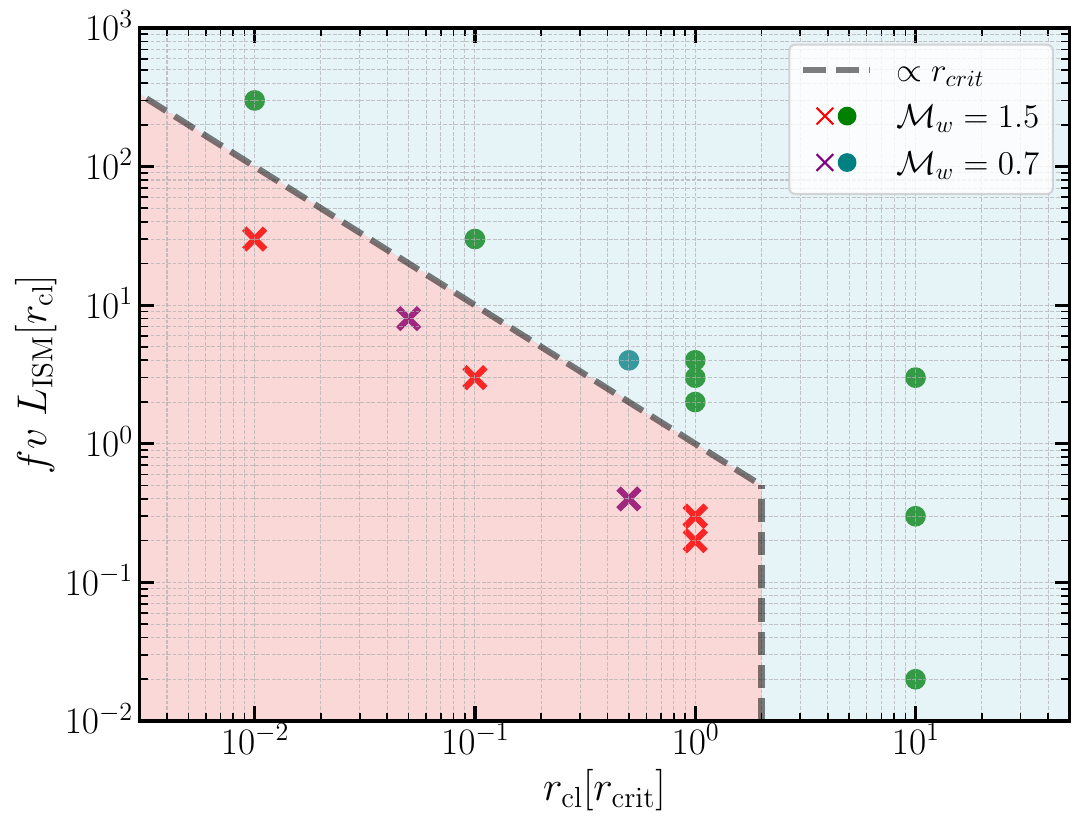}
    \caption{Emergence (dot) or absence (cross) of multiphase outflows as a function of cold mass-equivalent ISM depth, $f_v \lism$ (y-axis), and clump size $\rc$ (x-axis), expressed as fractions of the ISM clump size and the critical cloud radius, respectively. The dotted line indicates values proportional to the critical radius for single cloud survival.}
    \label{Results/fig:surv_dest}
\end{figure}

We compare this criterion with our results by extracting the effective ISM depth for our runs. In figure~\ref{Results/fig:surv_dest}, $f_v \lism$ is plotted as a function of the initial ISM coherent lengthscale. Survived runs are labelled as green (light blue) dots and destroyed as red (violet) crosses, respectively, for $\mathcal{M}\sim 1.5$ ($\mathcal{M}\sim 0.7$). As observed in section~\ref{sec:dynamical_ev}, ISM patches that are coherent over lengthscales larger than the single cloud criterion self-consistently survive regardless of $f_v$ or $\lism$. As we move towards the $\rc  <r_\mathrm{crit}$ regime, all runs that would classically experience destruction, now display survival above a certain $f_v \lism$ value. The threshold for survival follows a linear dependence. More specifically, we find it to follow $0.5 f_v \lism = \rcrit$ with a break-point at $\rc \approx r_\mathrm{crit}$, in excellent agreement with Eq.~\ref{eq:surv_criterion}.

Our criterion accurately extends to $\rc \ll \rcrit$. For instance, the left-most points in Fig.~\ref{Results/fig:surv_dest} lay two orders of magnitude below the traditional critical clump size. For both simulations we use a $f_v = 0.1$, with $\lism = 3000 \rc$ and $800$, where only the former, strictly satisfying the expected $ L >10^3 \rc$ limit for survival, can form a multiphase outflow. We further comment on this point in the discussion\S~\ref{sec:discussion:universal_criterion}.

\subsubsection{Mass distribution}
\label{Results/sec:mass_distribution}
\begin{figure}
    \centering
    \includegraphics[width=\columnwidth]{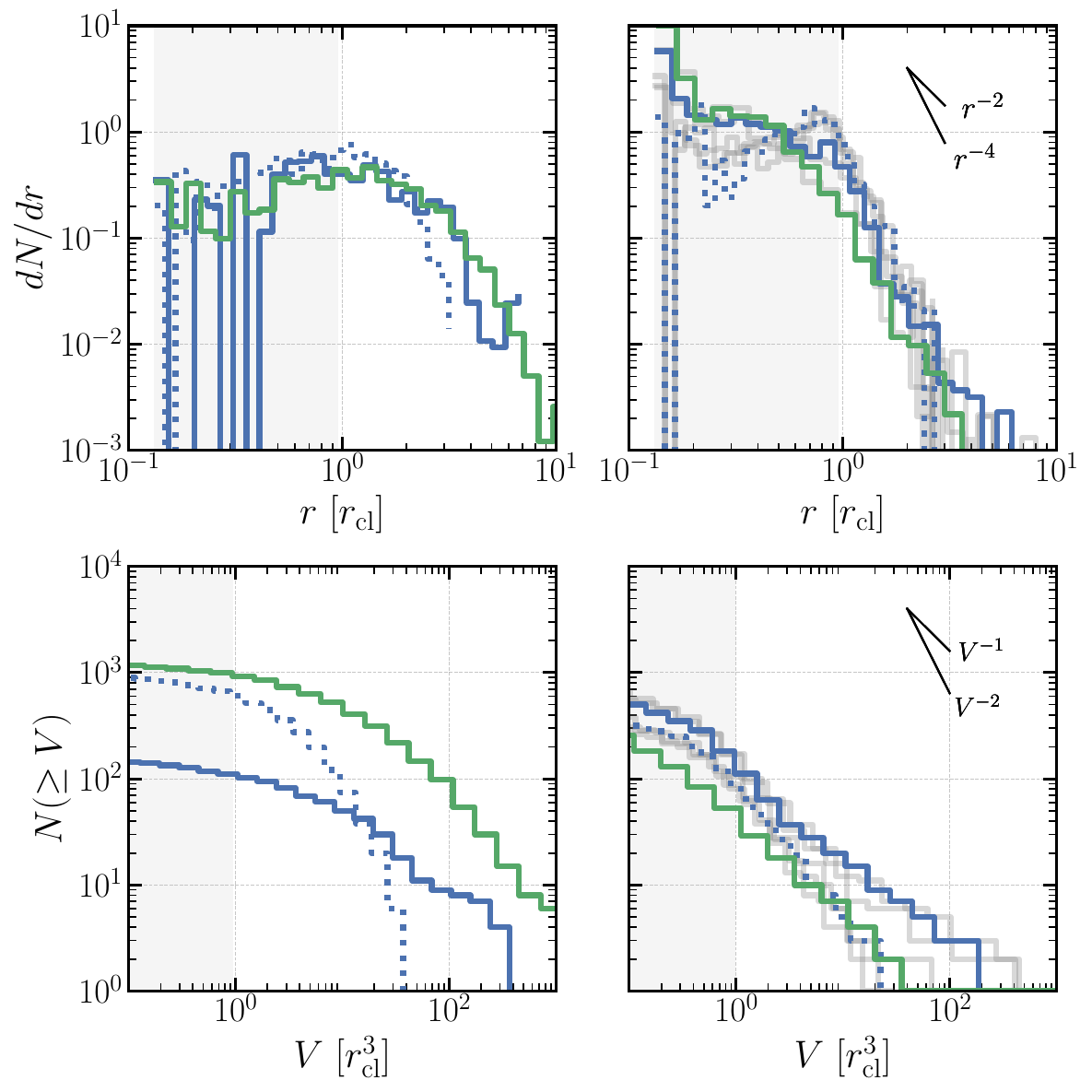}
    \caption{PDF (top) and CDF (bottom) for the clump size distribution of cold gas in the wind at time of initialisation (left) and at time of interaction with the wind (at $t \sim t_\mathrm{sh}$; right) for runs with $(f_v,\rc/\rcrit, \lism)=(10^{-1},1, 30\rc)$ and ($10^{-1}, 0.1, 300\rc$) as blue and green solid lines, respectively. The blue dashed curve shows a run identical to the solid blue but with $f_v=10^{-2}$, for completeness. Similar mass distributions are shown in light grey for later snapshots.}
    \label{Discussion/fig:clump_pdf}

\end{figure}
Figure~\ref{Discussion/fig:clump_pdf} shows the mass probability density function (PDF)  for the top row and cumulative distribution (CDF) for the bottom row of three runs, prior to the interaction with the wind (left panels) and during/after entrainment (right panels). We use \textsc{Scipy}\footnote{\url{https://docs.scipy.org/doc/scipy/}} union-find connected-component labeling (CCL) to identify clumps for each snapshot. Strikingly, we find that the cold gas distribution stabilises shortly after the wind-ISM interaction, and remains time-invariant throughout the simulation (shown in light gray).

The blue and green curves correspond to simulations with initial clump sizes of $\rc = \rcrit$ and $0.1  \rcrit$, respectively, with a volumetric filling factor of $f_v = 0.1$. Initially, both follow a similar bell-shaped density profile, reflecting our Gaussian-like initialisation of the ISM. These initial distributions do not reflect realistic ISM structures (see \S~\ref{methods:ism}), but are useful for isolating and testing survival parameters.
To highlight the impact of $f_v$ on the distribution of the clumps, we also show an identical run to $f_v = 10^{-2}$ in Fig.~\ref{Discussion/fig:clump_pdf} as a dotted blue curve.

For the second column, after $t\sim t_\mathrm{sh}$, the shape of the clump distribution changes significantly. The profiles across all three cases flatten into a power-law form consistent with $N(>V)\propto V^{-1}$, and $\dd N/\dd r \propto r^{-4}$, corresponding to a `Zipf like' mass distribution  $\dd N/ \dd m \propto m^{-2}$, in good agreement with previous results from turbulent, multiphase media \cite{gronke2022survival,das2024magnetic}, `shattering' simulations \citep{yaomandelker2025}, galactic winds \citep{tanfielding}, and ICM simulations \citep{li2015cooling,Fournier2024}. This shape is identically reproduced for the lower $f_v$ run, demonstrating that the $\propto m^{-2}$ mass distribution seems to be universally emergent -- and does not depend on the initial distribution. This power law distribution implies that  we do not find a clear characteristic, maximum, or minimum clump size accross the runs, and the powerlaw establishes around the initial mass range of the simulation (with the lower cut-off given by our resolution and the maximum cut-off given by the biggest clump (for the PDF) or total mass (for the CDF) of the system). 
Notably, runs with lower initial cold gas volume filling fractions $f_v$ produce smaller clumps in the wind.
This is more evident when comparing the end-tail of both blue curves in the CDF distributions, where the dotted line with $f_v = 10^{-2}$ terminates a factor of a few below the maximum clump size for $f_v = 10^{-1}$, a behaviour that we had already observed for the middle subfigure in Fig. ~\ref{Results/fig:multiplot}. 

Our findings show that regardless of initial ISM conditions, clump distributions during and after the initial interaction with the wind converge toward a universal clump mass distribution, effectively erasing any memory of the initial ISM configuration. This suggests that, at least structurally, multiphase outflows do not retain a direct imprint of the source galaxy's cold-phase morphology.

\subsection{Emergent turbulence in the  outflow}
\begin{figure}
    \centering
    \includegraphics[width=\columnwidth]{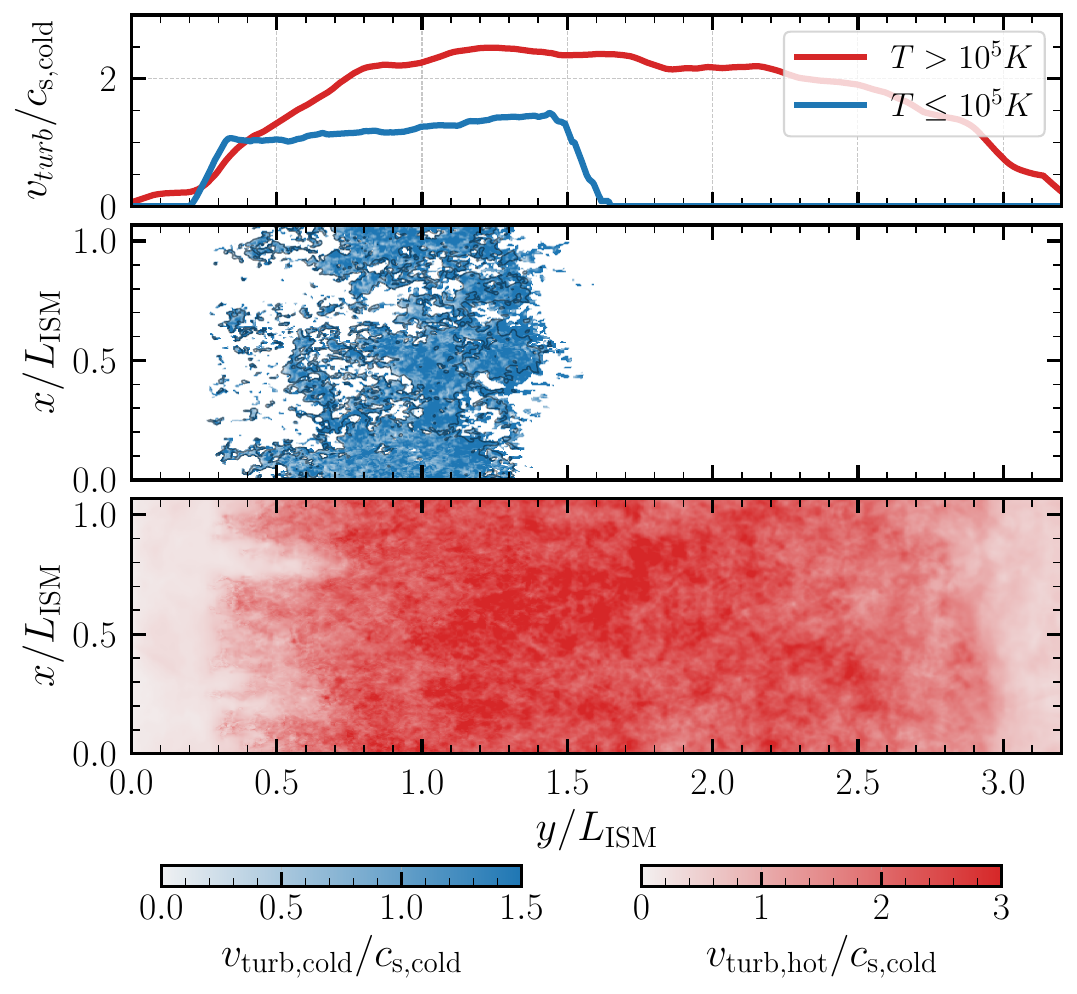}
    \caption{Averaged turbulent velocity of the cold and hot phases along the wind direction (top), and projected along the $z$-axis (middle and bottom), for the run $\rc/r_\mathrm{crit} = 10$, $f_v = 0.1$, and $\lism = 30\rc$ at $t \approx 0.2\, t_\mathrm{ent}$. We assume equipartition along the three dimensions when computing turbulent dispersion velocities.}
    \label{Discussion/fig:section_turb}

\end{figure}

As stellar and AGN-driven winds propagate through a porous ISM, the resulting interaction not only disturbs the cold gas but can also trigger turbulent motions from the initial hot gas laminar flow. Figure~\ref{Discussion/fig:section_turb} shows the mass-weighted projection of the velocity dispersion $v_\mathrm{turb} = (3/2\sigma^2_\mathrm{x} + 3/2\sigma^2_\mathrm{z})^{1/2}$, where $\sigma_{x/y}$ represents the standard deviation of the velocity component along the $x/z$  axis. The $3/2$ prefactor accounts
for unmeasured turbulent motions along the direction of the wind under the assumption of turbulence isotropy since, as we show below, turbulent speeds are well below $v_w$, effectively hiding $\sigma_y$ in the bulk flow of the outflow. $\sigma_z$ and $\sigma_x$ are self-similar (we further prove the self-similarity of turbulence in \S~\ref{sec:outflow_kinematics}).
Top panels show the marginalized spatial averages along the direction of the wind.

Figure~\ref{Discussion/fig:section_turb} shows that a volume filling, turbulent hot phase quickly develops throughout the simulation domain whereas cold gas (turbulence) remains more localised. 
Interestingly, while we have proven the bulk coupling of cold and hot speeds along the outflow direction in Fig,~\ref{Results/fig:mass_evol}, turbulence also shows to be coupled between the phases.  The hot, $10^6\, K$ gas only shows a factor of $\sim 2$ larger turbulent velocities than the cold phase, but both are of $\sigma\sim c_\mathrm{s,cold}$. While hot gas shows a larger dependence on the ISM depth and $f_v$, with lower average turbulence for lower cold gas fractions, cold gas turbulence consistently converges to $\sim c_\mathrm{s,cold}$ regardless of wind speed and ISM conditions.

\begin{figure}
    \centering
       \includegraphics[width=\columnwidth]{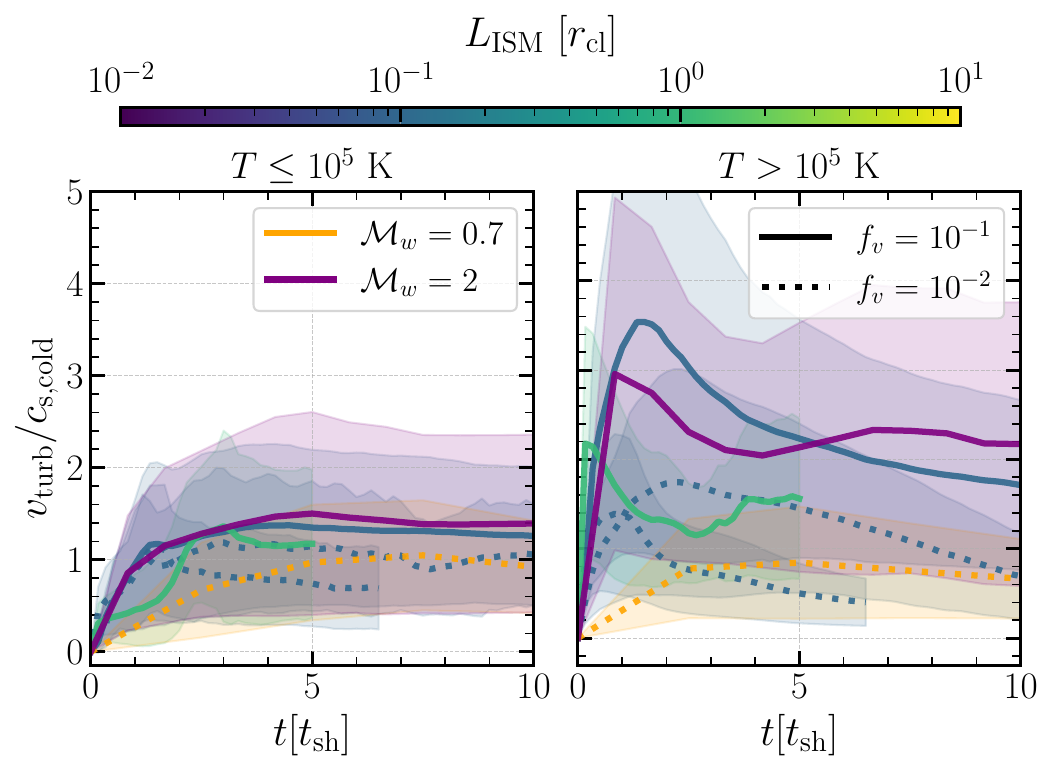}
    \caption{Average turbulent velocity (solid line) and 20-80 percentile ranges (shaded regions) as a function of time. The ISM depth is encoded in the colourbar. We represent different volume filling fractions with solid and dashed linestyles. Some addditional runs are added for comparison with wind Mach numbers $\mathcal{M}_w \approx 0.7,2$ in orange and purple, respectively.}
    \label{Discussion/fig:turbulence}

\end{figure}

Figure~\ref{Discussion/fig:turbulence} shows the time evolution of the mass-weighted mean turbulent velocity dispersion.
The left and right panels represent cases for the cold ($T \leq 10^5$ K) and hot ($T > 10^5$ K) phases, respectively. Linestyles follow those in Figure~\ref{Results/fig:mass_evol}, with curves colour-coded by the cold gas depth. For completeness, we show additional runs with wind Mach
numbers $\mathcal{M}_w = 0.7, 2$ in orange and purple, respectively.

We identify three key features:
\begin{enumerate}
\item All ISM configurations follow qualitatively a similar evolution. The peak in turbulence occurs around $t \sim t_\mathrm{sh}$, immediately at wind-crossing time, when shear between the hot wind and cold structures is strongest. 

\item The $10^4$ K gas consistently displays  turbulent motion of magnitude $c_\mathrm{s,cold}$, with peaks typically reaching values $v_\mathrm{turb} / c_\mathrm{s,cold} \sim 1.5$ and only occasionally exceeding it ($\sim 2 \, c_\mathrm{s,cold}$) for a handful of clumps as shown by the shaded regions, particularly for runs with narrower ISM depths. Turbulence for the cold phase remains stable at these values.

\item 
The hot gas retains a relatively laminar flow throughout most of the interaction with $v_w \gg v_\mathrm{turb, hot}$. Peaks in turbulent velocity are observed only at early times and lie an order of magnitude below $c_\mathrm{s,hot}$. Interestingly,
the peak in turbulence  shows a mild correlation with $f_v$ as opposed to the cold phase: higher volume filling fractions impose a larger inertial resistance to the wind flow and therefore a larger initial turbulence. All curves follow a gentle decay towards lower turbulent speeds, likely due to the merging of small-size clumps into larger structures.

\end{enumerate}

\subsection{Mass growth}
\begin{figure}
    \centering
    \includegraphics[width=\columnwidth]{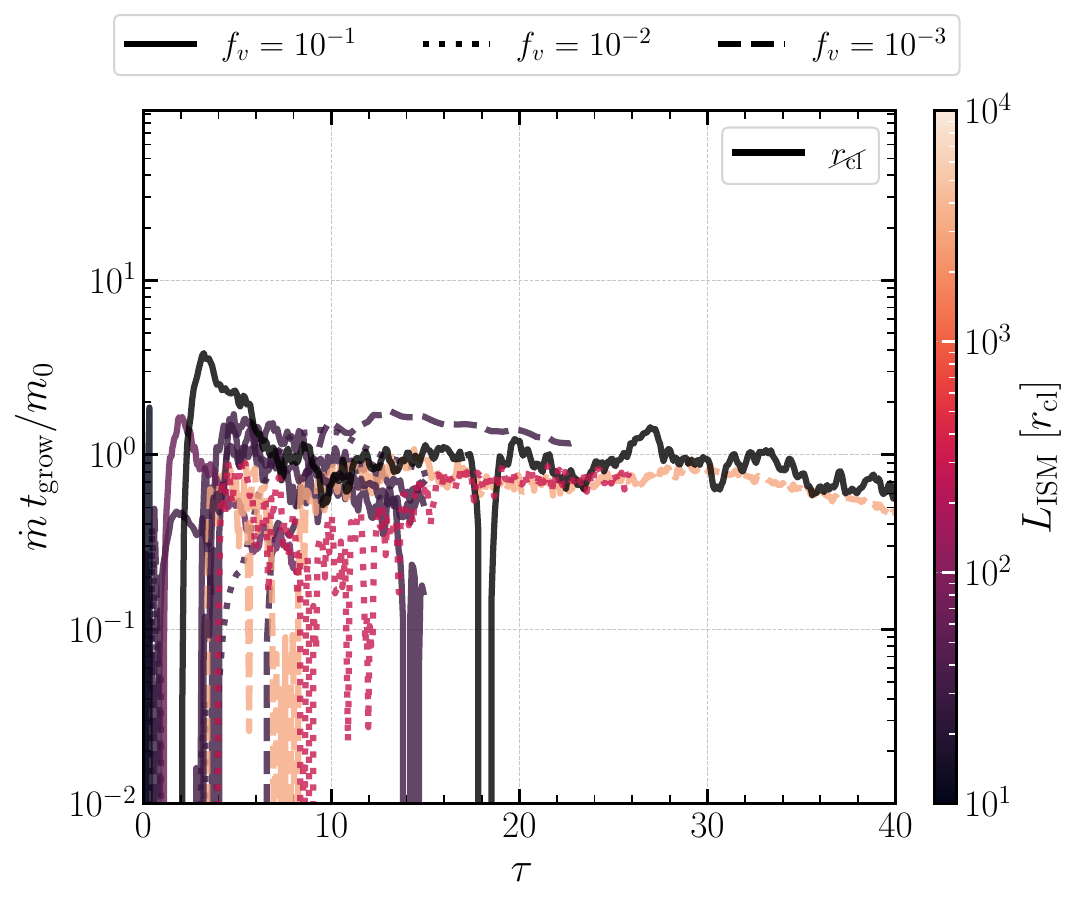}
    \caption{Mass growth of runs in figure~\ref{Results/fig:mass_evol}, divided by $m_0/t_\mathrm{grow}$, with $m_0$ as initial cold mass and $t_\mathrm{grow} = \chi(t_\mathrm{cool,mix}t_\mathrm{kh})^{1/2}$. These runs match those of Figure~\ref{Results/fig:mass_evol} and follow the same linestyle and colourcode.  In black, the mass growth of an additional scale-free ISM run with $(f_v,\lism) = (0.1,30)$.}
    \label{Discussion/fig:mdot}
\end{figure}

Previous work on the growth of cold mass in `wind tunnel'set-ups as well as in turbulent radiative mixing layers simulations \citep{gronke2020,tan2021,Fielding2020}.
\cite{tan2021} suggested that the effective cooling time of a turbulent, multiphase system is given by the geometric mean of the mixing and cooling time of the gas, thus, leading to a growth time of $t_\mathrm{grow}\equiv m/\dot m \sim \chi ( t_\mathrm{eddy} t_\mathrm{cool})^{1/2}$.
This was confirmed to hold in simulations of multiphase turbulence -- with an without magnetic fields \citep{gronke2022survival,das2024magnetic}.
Figure~\ref{Discussion/fig:mdot} shows the cold gas mass evolution normalized by this expected value. Our choice for $t_\mathrm{eddy}$ corresponds to the Kelvin–Helmholtz timescale of individual clumps, $t_\mathrm{kh} = \chi^{1/2} \rc / v_w$, and for $t_\mathrm{cool}$ we use the mixed gas cooling time $t_\mathrm{cool,mix}$\footnote{Note that this differs slightly from \citet{gronke2022survival} who use the mininum cooling time. The difference is, however, only a constant offset of a factor of $\sim 4$ and does not significantly change our findings.}.
The above-mentioned analytical expectation shows excellent agreement with the mass growth in our study,independent of volume filling fraction and global ISM properties.

Notice that we include a 'scale-free' ISM run for the mass growth analysis.  For this case, which contains a range of clump sizes, we evaluate the Kelvin-Helmholtz timescale using the largest clump in the sample, $t_\mathrm{kh,max}$. This `scale free' run is shown as the black curve in Fig.~\ref{Discussion/fig:mdot}. We further comment on this and other scale-free ISM runs in \S~\ref{sec:overdense_examples}.

\section{Discussion}

\subsection{A universal survival criterion for multiphase outflows }
\label{sec:discussion:universal_criterion}
Figure~\ref{Results/fig:surv_dest} shows that the survival of cold gas to the entrainment of a wind depends only on two parameters: the volume of the cold phase of a multiphase medium, $f_v$ and its depth in the direction of the wind propagation, $\lism$. The equivalent depth of the ISM, that is, the product of these variables, needs to be larger than the critical survival radius of a radiatively cooling plasma, which has been previously constrained by studies from \citet{gronke2018}. Specifically, survival is guaranteed for:
\begin{equation}
    f_v\lism \gtrsim 3 \, \text{pc} \, \frac{T_{\mathrm{cl},4}^{5/2}  \mathcal{M}_{\mathrm{wind}}}{ P_3 \Lambda_{\mathrm{mix}, -21.4}} \chi_{100} 
    \label{eq:rcrit_values}
\end{equation}
where $T_{\mathrm{cl},4} \equiv ( T_{\mathrm{cl}}/{10^4\, \mathrm{K}})$, 
$P_3 \equiv n T/ (10^3\, \mathrm{cm}^{-3}\, \mathrm{K}) $,
$\Lambda_{\mathrm{mix}, -21.4} \equiv \Lambda(T_{\mathrm{mix}})/(10^{-21.4}\, \text{erg}\text{cm}^3\text{s}^{-1})$, 
 $\mathcal{M}$ is the Mach number of the wind, and we write
$ v_{\mathrm{wind}} \sim c_{s,\mathrm{cl}} \mathcal{M} \chi^{1/2} $, and $\chi_\mathrm{100} = \chi / 100$.

Equivalent, since $f_v\lism \propto 1/n$ the column density of the ISM between the hot wind injection site and the surface needs to fulfill
\begin{equation}
    N_\mathrm{cold} \gtrsim 10^{18} \, \text{cm}^{-2} \, \mathcal{M} \frac{\chi}{100}  \left(\frac{\Lambda(T_4)/\Lambda(T_\mathrm{mix})}{0.1}\right)
    \label{eq:crit_density}
\end{equation}
in order for cold gas to survive in outflows.

\begin{figure}
    \centering
    \includegraphics[width=\columnwidth]{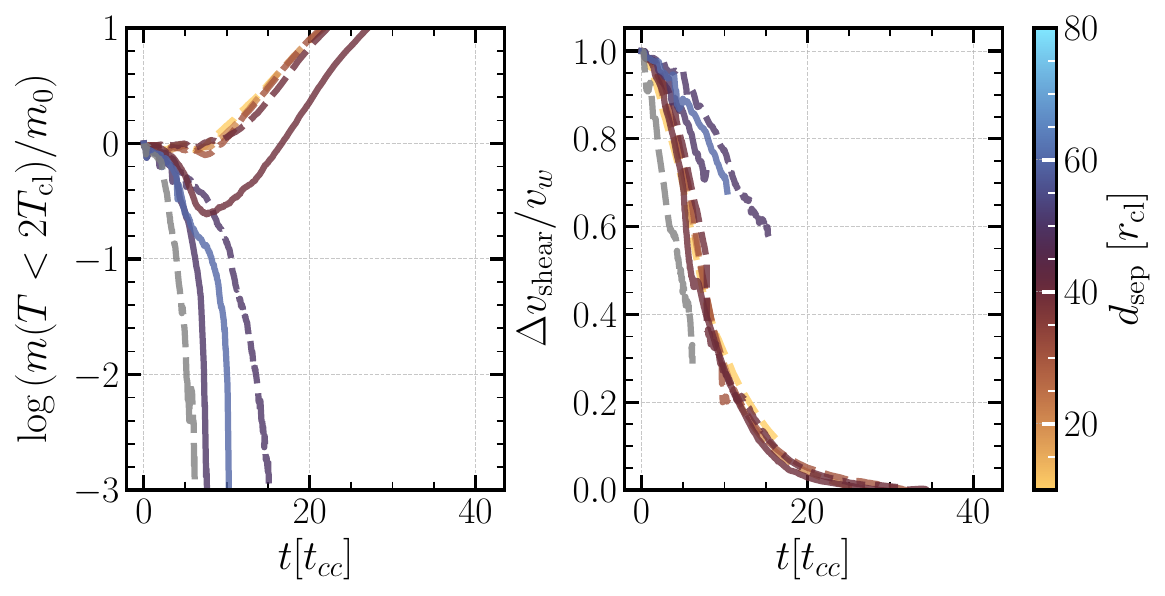}
    \caption{Mass evolution (left) and shear speed (right) for three aligned spheres along the wind direction, each with borderline destruction size $0.5\, r_\mathrm{crit}$ (dashed). A single isolated cloud does not survive (grey dashed). Simulations vary intercloud separation (colour bar) and we run and identical set-up varying cloud size to $0.75\, r_\mathrm{crit}$ (solid).}
    \label{Discussion/fig:cloud_separation}
\end{figure}

This means that the total projected cold gas along the direction of propagation of the wind requires densities above $\sim 10^{18}\text{cm}^2$ to efficiently entrain and grow cold mass within an outflow. Furthermore, the particular original geometry of the ISM is irrelevant to the question of survival: as long as the column density is met, a cold outflow will be formed. Consequently, this criterion applies more generally to any scenario where cold gas is accelerated by a more tenuous phase.
Importantly, here $L_{\rm ISM}$ does not represent the entire heigh / depth of the ISM but rather the distance from the hot medium injection (e.g., the clustered SN site) to the disk surface.

These survival criteria are directly analogous to the ones found in the single-cloud case \citep{gronke2018}. Note, however, that there is ongoing debate in the literature about the exact formulation of this criterion 
\citep{LiHopkins2020,Sparre2020,Kanjilal2020,abruzzo2022simple} with divergent criteria found for $\chi \gtrsim 10^4$. As we focus in this study on $\chi\approx 100$ (see also \S~\ref{sec:overdense_examples} where we check the robustness of the survival criterion  for $\chi=1000$) 
where the criteria mostly agree, we do not contribute to the discussion of survival timescales. We rather note that independent of the exact formulation of the `single cloud criterion', there is a generalised survival criterion applicable to complex ISM morphologies, captured by the effective depth of the ISM patch and where the details enter in the form of an $\rcrit$. 

It is possible to think of a limit in volume filling fraction under which our survival criterion Eq.~\ref{eq:surv_criterion} will break apart. The underlying assumption in our derivation is that separate clumps will always interact among each other in a multicloud set-up. We expect a certain threshold in $f_v$ for which inter-cloud distances will be larger than their `radius of interaction', and this assumption will no longer hold. We find no evidence for the existence of this break-down at $f_v = 10^{-3}$, the lower limit in the parameter space of our suite of simulations. Exploring further this parameter space is computationally expensive, so we examine the limiting radius of influence for the ideal-case scenario and extrapolate to ISM conditions.

The top panels of Fig.~\ref{Discussion/fig:cloud_separation} show the mass (left) and relative velocity evolution (right) of 3 spherical clouds aligned along the direction of the incoming wind, in analogy to our multicloud analysis, now color-coded by the separation between them. We choose the individual clouds sizes to be $\rc/\rcrit \approx 0.5 < 1$, i.e., not to satisfy the survival criterion for a single cloud. We show the destruction status of a single cloud as the grey dashed line in Fig.~\ref{Discussion/fig:cloud_separation}. The system does however, experience cold gas growth for cloud-cloud separations below $\sim 40 \rc = 4 \chi^{1/2}\rc$, but destruction reemerges as clouds are initialised with distances larger than this limit. From our isotropic ISM distributions where $f_v = \rc^3/d^3$, this break down in survival occurs for $f_v \approx 10^{-5}$ (see Appendix~\ref{app:resolution}).
This critical cloud separation aligns with previous single cloud crushing studies who found that the extend of single-cloud tails before destruction is $\sim 2\chi^{1/2} \rc$ \citep{gronke2020}. This implies that survival is also dependent on a second requirement, $4f_v^{1/3}\chi^{1/2} \gtrsim 1$, criterion often met for ISM gas conditions. 

Note that this approach of considering the shielding effect between individual clumps is similar to that of previous `multi cloud crushing' studies \citep{pittard2005dynamical,seidl2025,forbes2019hydrodynamic}.
Recently, \citet{seidl2025} systematically studied a variety of multi-cloud set-ups with the inclusion of radiative cooling  and obtained a survival criterion defined from the overlap of what they define the `effective volume' of single clouds. They define an effective cloud volume approximated as a cone aligned with the wind, with dimensions $L_\parallel = a_{\parallel}\rcrit$ and $L_\perp = a_{\perp} \rcrit$ with $(a_\parallel,\,a_\perp) = (7.5,\,3)$, and use this to derive a critical effective filling fraction, $\tilde f_\mathrm{v,crit} \approx 0.24$ above which they find cold gas survival. 
Translating to our set-up, the picture of overlapping volumes can lead to two scenarios: one where the intercloud separation $d_\mathrm{cl} > L_\parallel$, and one where $d_\mathrm{cl} < L_\parallel$\footnote{Because $d_\parallel \approx d_\perp$ in our simulations, we take $L_\parallel$ as the main constraining volume length, since distances in the direction of the wind are the most stringent for the survival of cold gas, see \S~\ref{sec:discussion:universal_criterion}}. The former case, $L_\parallel > d_\mathrm{cl}$, generally holds for our runs with $\rc < \rcrit$ and $f_v \geq 10^{-3}$. We can approximate for this case the effective volume filling fraction $\tilde f_{\rm V}$ by considering an elongated volume $\tilde V = A (L_{\rm ISM} + a_\parallel)$ where $A=L_z L_x$ is the cross section of our domain. Thus,
\begin{equation}
    \tilde f_{\rm V} \approx \frac{N\rc^3}{\tilde V} = \frac{f_v}{1 + a_\parallel(\rcrit/\lism)}
    \label{eq:fv_Seidl25}
\end{equation}
where $N$ is as above the number of clumps.

We can extract an analogous effective critical folume filling fraction $\tilde f_\mathrm{v,crit}$ by substituting our survival criterion $f_v\lism = \rcrit$ into Eq.~\eqref{eq:fv_Seidl25}, $\tilde f_\mathrm{v,crit} \approx 1/(a_\parallel + f_v^{-1} )$.
This shows that, for $L_\parallel > d_\mathrm{cl}$ (i.e. $\rc \ll \rcrit$ and $4f_v^{1/3}\chi^{1/2} \geq 1$), the effective critical filling fraction should in fact vary with $f_v$, not remain at a fixed value. However, \citet{seidl2025} only explored volume filling fractions of the order of $f_V\gtrsim 10^{-1}$, for which our results do align well with their findings (e.g. for $f_v = 0.3$, $\tilde f_\mathrm{v,crit}\approx 0.1$).  \\

In summary, our generalised survival criterion provides a simple yet powerful framework for predicting when multiphase gas can persist in galactic outflows. It shows that the emergence of cold gas in winds is not determined by the details of individual clouds, but rather by the integrated cold gas column between the wind-launching region and the surrounding medium. This criterion not only unifies a range of previous single- and multi-cloud results, but also explains why cold material is ubiquitous in observed galactic winds: as long as the effective column exceeds the critical threshold, cold gas will survive, grow, and assemble into extended shells or plumes irrespective of the original ISM geometry.  

Beyond galactic winds, these results can have implications for simulations of ram-pressure stripping in galaxies moving through the intracluster medium (ICM; \citealp[e.g.][]{ritali2025}). There, our criterion suggests the fate of stripped gas -- and whether it forms a multiphase tail or is rapidly destroyed and mixed -- likewise depends on the effective cold gas depth along the flow direction. 
This prediction can be tested with high-resolution simulations that systematically vary ISM structure and distribution.

\subsection{The effect of scale-free ISM morphologies, larger overdensities and different wind speeds}
\label{sec:overdense_examples}
\begin{figure}
    \centering
    \includegraphics[width=\columnwidth]{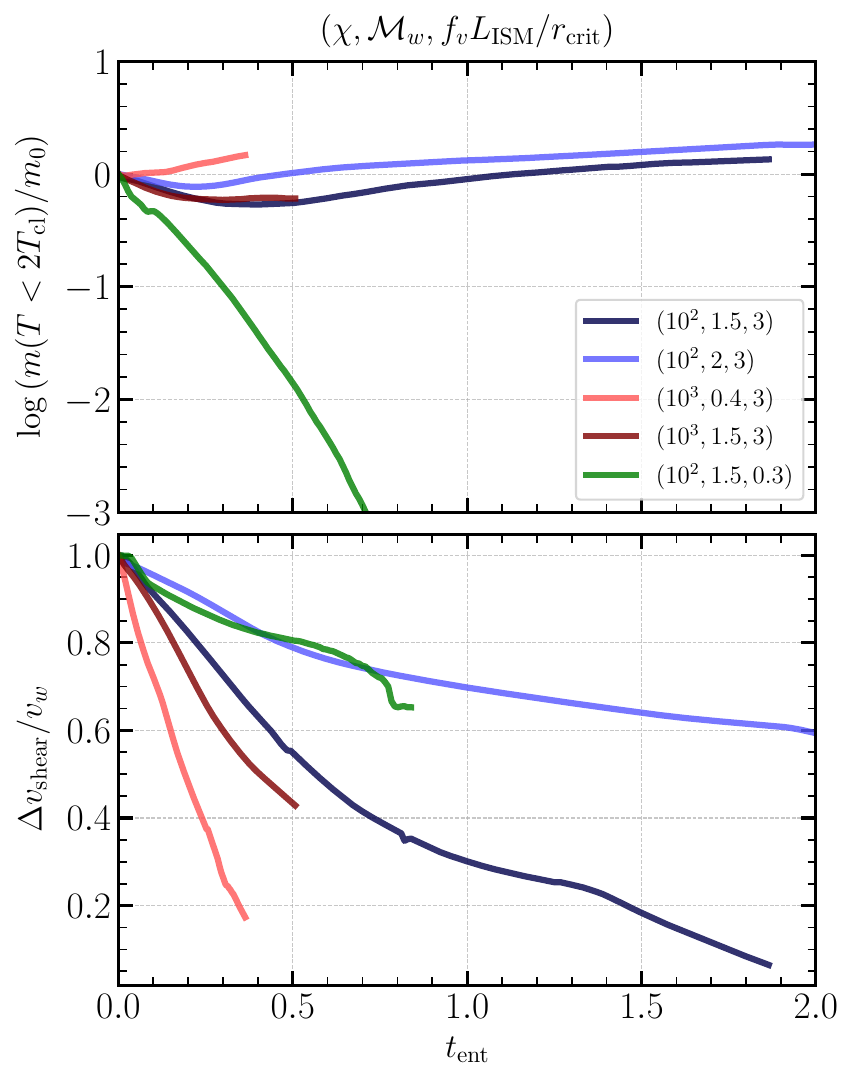}
    \caption{Mass and shear evolution for `scale-free' ISM runs with $f_v\lism = 3\rcrit$ and one with $f_v\lism = 0.3\rcrit$ (green). Dark blue: clumps from $0.05\,r_\mathrm{crit}$ to $r_\mathrm{crit}$. Purple: same as blue with with Mach number $\mathcal{M}_\mathrm{w} = 0.7 \, ( 440 \text{kms}^{-1} )$. Light red: 'scale-free' ISM with overdensities $\chi = 10^3$. Dark red: same as light red with Mach number $\mathcal{M}_\mathrm{w} = 1.5\, ( 725 \text{kms}^{-1} )$.}
    \label{Discussion/fig:overdense_evolution}

\end{figure}
Observations and simulations of the multiphase interstellar medium \citep{federrath2009fractal,elmegreen2004} reveal gas structure extending spatially over several orders of magnitude. This is different to the ISM density field generated in the prevous sections where  cold clumps possess a characteristic scale. This raises the question of how our survival criterion connects to a realistic, scale-free ISM structure.

The dark blue line in Fig.~\ref{Discussion/fig:overdense_evolution} shows the evolution for a simulation where we initialise the ISM for a range of clump sizes, spanning over an initial $\rc \sim 10r_\mathrm{crit} = 80\, d_\mathrm{cell}$ to $\rc \sim\, r_\mathrm{crit}= 8\, d_\mathrm{cell}$ using the ISM generation algorithm described in \S~\ref{methods:ism} (the range of cloud sizes is shown in the appendix figure~\ref{fig:scale_free_cdf}). We use an effective cold gas depth of $f_v\lism = 3\rcrit$ with a volume filling fraction of $f_v\approx 0.15$. As shown for the blue line in the top panel of Fig.~\ref{Discussion/fig:overdense_evolution} the cold gas is well-entrained within $t_\mathrm{ent}$, growing above its initial mass. This behaviour follows the expectation from our analysis. Since the emergent survival criterion in Eq~\eqref{eq:crit_density} reveals survival does not depend on an intrinsic lengthscale of the ISM, it equally applies here for our scale-free ISMs. We further prove this point by analysing the evolution, in solid green, of an identical set-up but with $f_v\lism \approx 0.3 \rcrit$, where the cold phase is completely absent by $t_\mathrm{ent}$.

To explore multiphase outflow formation at larger overdensities, we conduct additional runs with higher wind temperature. All runs in Fig.~\ref{Discussion/fig:overdense_evolution} use $f_v\lism = 3\rcrit$ and $f_v\approx 0.15$.
The purple curve shows a scale-free ISM run with the original parameters: $\chi = 100$, $T_w = 10^6$ K, $T_\mathrm{cold} = 10^4$ K, and $\mathcal{M}_w = 2$. The light red and dark red curves use an increased wind temperature of $T_w = 10^7$ K (giving $\chi = 10^3$ with $T_\mathrm{cold} = 10^4$ K), with $\mathcal{M}_w = 0.4$ ($v_w \approx 220$ \kms) and $\mathcal{M}_w = 1.5$, respectively. Despite spanning different temperature regimes and parameter combinations, these runs are initialised to satisfy Eq.~\eqref{eq:surv_criterion}. All cases show clear survival and accurately follow the survival criterion.

\subsection{Multiphase morphology}
\label{sec:fv_fa}

\begin{figure}
    \centering
    \includegraphics[width=\columnwidth]{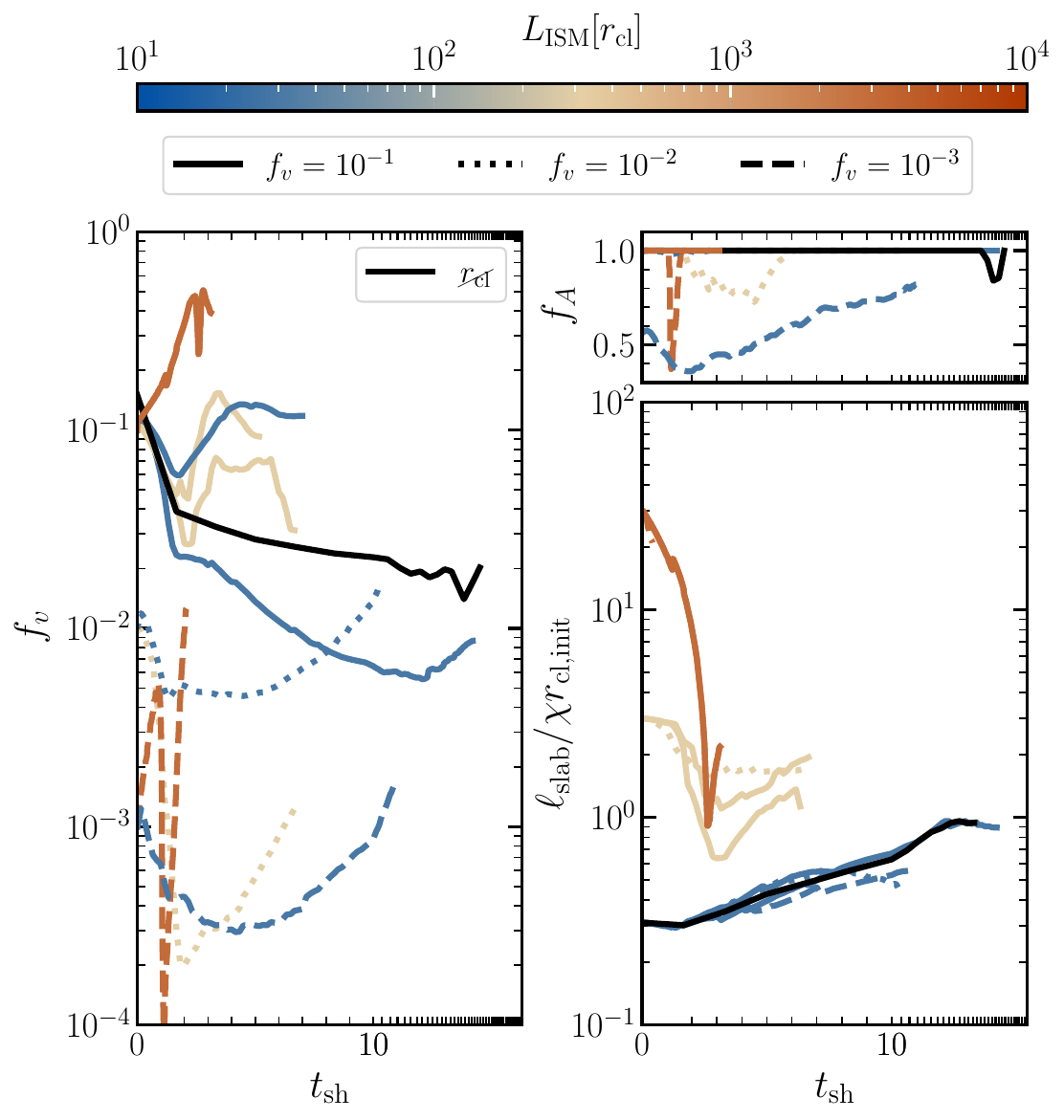}
    \caption{Volume filling fraction (left), areal covering fraction (top-right) and cold gas shell extent in the wind direction (bottom-right) evolution for runs of different initial clump size (colourbar). We express $\ell_\mathrm{slab}$ as a fraction of a cloud's tail, $\sim \chi \rc$. The initial volume filling fraction is denoted by the linestyle.}

    \label{Discussion/fig:fv_fa}
\end{figure}

The ubiquity of cold outflows at both low and high redshifts in turn requires a launch mechanism that can efficiently expel large numbers of cold clumps \citep{veilleux2005galactic,veilleux2020cool,thompson2024}. 
Similarly, observations of galactic outflows show that while cool gas permeate the field of view with covering fractions near unity, their inferred volumetric fraction are well below unity, $f_v \ll 1$ \citep*[see][for more references]{xu2022classy, xu2023classy, thompson2024}. 

Fig.~\ref{Results/fig:multiplot} demonstrates that cold gas preferentially concentrates in 'shells' where $f_A\sim 1$ in runs exhibiting cloud survival.
In Fig.~\ref{Discussion/fig:fv_fa} we explicitly show the cold gas volume filling fraction (left panel), its area covering fraction (top right panel), and the extent of this cold gas shells along the direction of the wind (bottom right panel) in runs with $f_v \lism > \rcrit$, i.e., in which multiphase outflows are formed. We colour-code the curves by ISM depth and the linestyles denote the initial volume filling fraction. We compute both $f_v$ and $f_A$ within the limits of the cold outflow, i.e. from the negative- to positive-most cold gas component along the $y$ axis, and within the full $x, z$ extent. This allow us to set an upper-limit on both parameters. The total $f_v$ is the ratio of number of cold gas cells to total within these limits. For this same domain we compute the 
covering fraction $f_A$ from the left-most boundary in the direction perpendicular to the wind velocity, i.e., along the $z$ direction.

The volumetric filling fraction for these runs remains well below unity at all times. Despite initial drops, $f_v$ consistently increases with time for all survival scenarios. Meanwhile, the areal covering fraction of winds, $f_A$, asymptotically reaches unity. The general picture shows that $f_A \sim 1$ throughout; remarkably, even for initial ISM conditions with extremely low volume filling fractions, the covering fraction naturally tends toward unity, as in the other cases. This is expected, as the clumps gather around $\rcrit$ in order to survive, occupying larger projected areas than their original sizes. 

Observations from \cite{xu2022classy} of line multiplets and doublets in 45 low-redshift starburst galaxies showed that outflow winds retain a large covering fraction of approximately $f_A \sim 0.64$, in very good agreement with our results, where we find $f_A \sim 1$. On the other hand, \cite{xu2023classy} estimated the volumetric filling fractions of cold gas in M82 outflows from the resolved $[\text{S II}], \lambda\lambda6717, 6731$ emission lines, and found very low values of $f_v \sim 10^{-3} - 10^{-4}$. While these values satisfy $f_v \ll 1$ as in our simulations, they are 1-2 orders of magnitude lower than most of our runs. Furthermore, our simulations show $f_v$ increasing monotonically  (cf. ~\ref{Discussion/fig:fv_fa}), whereas their estimates show $f_v$ decreasing with radius, with a dependence $f_v \propto r^{-1.8}$. This discrepancy is, however, an expected consequence of our set-up, as wind-tunnel simulations miss the expanding nature of outflows in which cold gas redistributes across larger volumes as it travels with the wind. Pressure and density properties also evolve in such scenarios, leading to a decrease of the mass transfer rate between the hot and cold material \citep{Dutta2025}. We also caution that our estimates of $f_v$ and $f_A$ are restricted to a shell of gas and, thus, are strictly speaking upper limits compared to the observed values.

Note that in Fig.~\ref{Discussion/fig:fv_fa}  some runs show an apparent plateau in volume filling fractions of $f_v\sim 10^{-1}$. This could be attributed to artifacts of the periodic boundary conditions, which can alter the mass growth once gas shells increase in mass. Figure~\ref{Discussion/fig:mdot}, however, shows that the mass growth of the same runs remain close to its predicted value.
Indeed, it is possible that this property is inherent to multiphase systems which can undergo a strong fragmentation process, and hence reduce the overall volume the cold gas is filling. This could explain why mass growth continues despite a fixed cold gas volume fraction.

While the cold gas volume filling fraction mostly remains low, some of the projections in Fig.~\ref{Results/fig:multiplot} clearly  show the formation of ``cold gas shells'', i.e., confined regions of the outflows where most of the cold material is concentrated, leading large $f_
v$ locally. In the bottom-right panel of Fig.~\ref{Discussion/fig:fv_fa}, we analyse the temporal extent of cold gas along the wind direction, $\ell_\mathrm{slab}$. The initial expansion or compression of the ISM is tightly coupled with its initial depth: systems with $\lism > 10^2\rc$ are compressed, in contrast to runs with $\lism \leq 10^2\rc$, which expand.

This can be understood by equating the shear timescale of the wind $t_{\rm sh}\sim \lism/ v_w$ with the reaction time of individual clumps $t_\mathrm{ent}\sim \chi \rc / v_w$, which shows that this transition first occurs at $\lism \sim \chi r_\mathrm{cl, min}$. In other words, initial depths shallower than the expanding tail of surviving clouds lead to an apparent extension of the cold shell whereas is the initial distribution is smaller, the cold gas extends to $\sim \chi \rc$.

Not only do single cloud tails determine whether the cold expands or contracts initially but also set a minimum evolving shell size for multiphase outflows. This is shown by the curves in the bottom right of Fig.~\ref{Discussion/fig:fv_fa}, where all the curves settle close to $\ell_{\rm slab} \sim \chi \rc$, after which further fragmentation and mass accretion can lead to slow and continuous increase in size.
The compression of the interstellar medium (ISM) is especially important when examining outflow energetics. In particular, in studies of momentum- and energy-driven outflows, the presence of a compact shell of cold gas may cause a momentum-driven outflow to resemble an energy-driven phase. This occurs because the hot gas performs $P\dd V$ work on the dense cold shell, rather than escaping through a porous medium, a phenomenon common in AGN driven winds \citep{faucher2012physics,ward2024}.

This criterion for the expansion and compression of cold gas could explain the formation of both plumes and shells observed in starburst galaxies. For example, $3.3,\mu$m PAH emission in \citep{fisher2025jwst} reveals plumes extending up to 200–300 pc where the cold dust appears to align with the direction of the wind and is embedded with clumps of sizes 5 - 15 pc. These observations are consistent with our findings, where for clumps sizes of 5 pc in a sufficiently narrow ISM above the SNe event ( $\lism < \chi \rc \approx 500$pc), cold gas should form plumes that extend by 2 orders of magnitude with respect to their initial size.
Similarly, studies by \citet{ha2025rupke, rupke2019} on the Makani galaxy and by  \citet{lopez2025} on M82 focus on the formation and properties of outflows traced by O IV and H$\alpha$ emission shells. These oberved cold gas morphologies stand in stark contrast to the cometary structures predicted by single-cloud simulations \citep[see discussion in][]{thompson2024}. 
For example, \cite{lopez2025} shows the formation of slabs and arcs of cold gas in M82 that are 14–50 pc deep along the direction of the wind. If one compares this to the values of e.g., the red curve in the bottom right panel of Figure~\ref{Discussion/fig:fv_fa}, we find that for intial clump sizes of $\rc \sim 0.1 \rcrit \sim 0.2$ pc (using the fiducial values in Eq.~\eqref{eq:rcrit_values}), the final shell size reaches $\sim$10 pc, in agreement with their findings. Another critical point against single-cloud studies is their inability to form arc-like structures. In our case, some of the morphology maps (middle panel of Figure~\ref{Results/fig:multiplot}) begin to show asymmetric arc-like features. Although not as clear as in the M82 observations \citep{thompson2024,lopez2025}, the expanding background absent in our simulations is likely essential to accentuate these structures. Previous single-cloud studies also highlight the potential role of magnetic fields in forming more filamentary structures \citep{shin2008magnetohydrodynamics,Gronnow2018,hidalgo2024better}. While we do not include magnetic fields in this work, we hope to explore it in the future.

Another question we  address  is how the multiphase morphology of the emergent wind is related to the morphology of the originating ISM. In other words, do galactic winds retain any imprint of the initial cold gas scale? Our results show that the mass distribution in an outflow is governed by Zipf’s law ($\dd N/\dd m\propto m^{-2}$; cf. Fig.~\ref{Discussion/fig:clump_pdf}). This distribution has been observed in various simulation setups:  galactic outflows \citep{tanfielding}, multiphase turbulence \citep{das2024magnetic,gronke2022survival}, and simulations of the ICM \citep{LiMODELINGACTIVE2014a,Fournier2024}.
In our study of ISM–wind interactions, this power law permeates the cold gas structure across time, demonstrating that even with different initial conditions, the cold phase rapidly converges toward this behaviour, and suggesting that the power law is a universal attractor for gas arising from cold–hot phase interactions -- and the ISM structure is not imprinted in the winds.
Observations of nearby multiphase winds reach now the resolution required in order to compare this clump mass distribution to data \citep[e.g.][]{lopez2025,fisher2025jwst}. However, a quantitative analysis is still outstanding.

\subsection{Outflow kinematics}
\label{sec:outflow_kinematics}
Multicloud setups offer a closer approximation to realistic galactic outflows, as shown above by several properties of ISM–wind interactions not captured in single-cloud studies.
For instance, the mass-carrying phase of a single outflow can exhibit a wide range of bulk speeds along the wind direction (see Fig.~\ref{Results/fig:vel_phase}). 
Specifically, we predict a modified `Schechter-like' function for the cold gas spread in mass and velocities in winds (cf. \S~\ref{sec:phase_plot}) which can be in principle observable.
We show that the characteristic $v/v_c\exp(-(v/v_c)^2)$-distribution forms after $t\sim t_\mathrm{sh}$, while during the wind-ISM initial interaction, the clump distribution is semi-random, with most of its material travelling at $v_c \sim 0$. The peak and normalisation increases with time, until reaching $v_c \approx v_w$ at time $t \sim t_\mathrm{ent}$, i.e. typically after tens of Myr.

This implies that using, e.g., `down the barrel' observations and comparing low-ionization absorption line shapes one can infer at which evolutionary stage an observed outflow is and what this implies for the distance the cold gas has travelled, allowing e.g. to estimate the mass outflow rate of the wind.
In fact, observations of nearby galaxies have shown potentially compatible functional forms in their low-ionization species absorption lines reaching $\sim$ hundreds of kilometers per second \citep{Rivera-thorsen2014, barger2016down}. 
Further observations and comparisons to simulated outflow profiles such as the ones suggested here will help constrain the nature of galactic winds.

\begin{figure*}
    \centering
    \includegraphics[width=\textwidth]{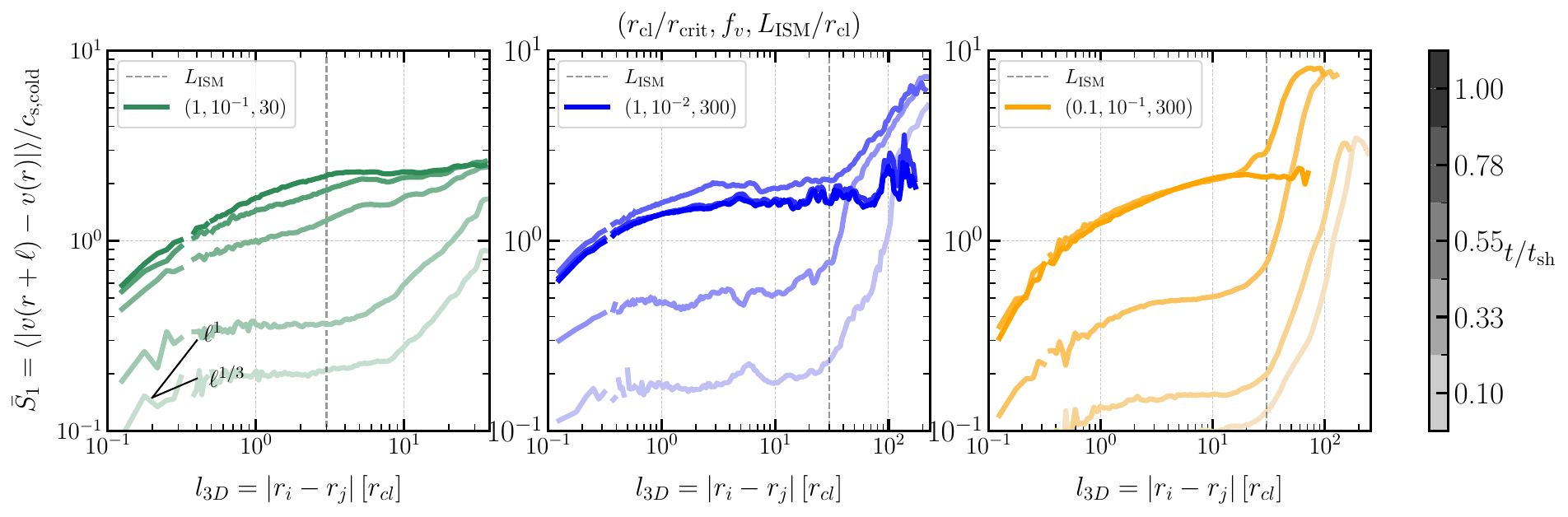}
    \caption{
    First-order structure functions for the cold phase of the wind ($T<10^5\,$K), shown for three simulations (from left to right panel): $(\rc/r_\mathrm{crit}, f_v, \lism) = (1, 0.1, 30\rc)$, $(1, 0.1, 300\rc)$, and $(0.1, 0.1, 300\rc)$. Temporal evolution is indicated by line colour saturation in units of the shear time $t_\mathrm{sh}=\lism/v_{\rm w}$. Due to our resolution, clump pairs are undetected at certain separations, leading to gaps in the VSF where the contribution is 0.}
    \label{Discussion/fig:vsf}

\end{figure*}

We also study the emergent turbulent motion of the (initially laminar) outflow (\S~\ref{Discussion/fig:section_turb}). There, we show that the phases remain kinematically coupled throughout evolution. In particular, the turbulent velocity of the cold phase converges to $\sim c_\mathrm{s,cold}$ across all simulations. The turbulent velocity of the hot phase, on the other hand, is slightly higher -- about a factor of $\sim 2$ larger than that of the cold phase -- and shows a weak dependence on the initial volume filling factor $f_v$. Specifically, the hot-phase turbulent velocity increases by roughly a factor of $\sim 2$ when $f_v$ is increases from $10^{-2}$ to $10^{-1}$, indicating that higher cold-gas volume fractions lead to stronger turbulent motions in the hot medium. We further investigate the turbulent cascade in these outflows by computing the velocity structure function (VSF hereinafter), a standard diagnostic of (multiphase) turbulence \citep{vonHoerner1951,li2020turbulence,Fournier2024}.

Figure~\ref{Discussion/fig:vsf} shows the 
first-order velocity structure function of cold gas for runs with $(\rc/r_\mathrm{crit}, f_v,  \lism/\rc) = (1, 10^{-1}, 30)$ (left), $(1, 10^{-2}, 300)$ (middle), and $(0.1, 10^{-1}, 300)$ (right), i.e., runs in which the cold gas survives. The line colour saturation towards darker tones indicates the temporal evolution of the VSF as the wind traverses $\lism$, as indicated by the colour-bar.

For these three cases, the VSF progressively flattens as the ISM is traversed by the wind, reaching a plateau at $t \approx t_\mathrm{sh}$, when the wind has traversed the ISM, displayed by the darkest solid line in each panel. At the smallest scales, the curves display a Kolmogorov relation following a $\ell^{1/3}$ power law. This relation is absent at early times, but once the wind has traversed the full depth of the cold gas, the inertial turbulent regime develops, spatially spanning nearly two orders of magnitude from $\rc$ to $\lism$ for the run shown in the right panel of Fig.~\ref{Discussion/fig:vsf} ($f_v =0.1$, $\rc / \rcrit = 0.1$, $\lism = 300 \rc$). This injection scale can be inferred from the flattening of the VSF curves at large scales, which consistently matches $\ell \sim \lism$ for all runs. 
The temporal evolution of the VSF magnitude is in agreement with Fig.~\ref{Discussion/fig:turbulence}, where we showed turbulence peaks for both cold and hot phases at $t\sim \lism / v_w$ reaching $v_\mathrm{turb} \sim c_\mathrm{s,cold}$.
By comparing the first and second panel of Fig.~\ref{Discussion/fig:vsf}, one can see that $f_v$ does not alter the inertial regime of the VSF where Kolmogorov's turbulence applies; in both cases the driving scale $\sim \lism$. In addition, the magnitude is $S_1(\lism)\sim c_{\rm s,cold}$. This behaviour is also shown in Figure~\ref{Discussion/fig:turbulence} where $f_v$ has negigible impact on the cold gas rms velocity.

We are not aware of studies of the VSF in the cold gas component of star-driven outflows in galaxies. Existing work has been limited to cool-core clusters, where AGN activity produces long H$\alpha$ filaments \citep{li2020turbulence,hu2022signature}. In cases where outflows are supersonic, a short $t_\mathrm{cool}$ leads to partial thermalisation of the energy cascade, steepening the scaling to $\ell^{1/2}$. This is the case for Perseus, as shown by \cite{li2020turbulence}, where the effect was first identified, and later confirmed through simulations in \cite{hu2022signature}. In the latter study, simulations with subsonic outflows recover the $\ell^{1/3}$ scaling, consistent with our results, and show a similar temporal trend in the VSF, an increase with time that eventually settles into the Kolmogorov cascade after wind entrainment.
Our results suggest that turbulence does develop naturally in multiphase winds, and that measuring the VSF can reveal not only the evolutionary state of the wind but also the driving scale which in turn can constrain the original ISM from which the outflow emerged.

\subsection{Comparison to previous work} 
\label{sec:discussion/comparison_sims}
A large body of literature has focused on cold gas-wind interactions. The simplest cases are, as already preluded in \S~\ref{sec:intro}, the single `cloud crushing' studies \citep{klein1994,schneider2016couplingmassmomentum,scannapieco2015} which have been extended with radiative cooling to develop a cold gas survival criterion \citep{gronke2018,LiHopkins2020, Kanjilal2020} -- which our general criterion is consistent with (see \S~\ref{sec:discussion:universal_criterion}).

Several studies also focused on ensemble of clouds and the associated `shielding' effects \citep[e.g.][]{Poludnenko2002,Aluzas2012}. As mentioned in \S~\ref{sec:discussion:universal_criterion}, specifically, \citet{seidl2025} performed multi-cloud simulations that include radiative cooling which allowed them to derived a critical volume filling fraction from the overlapping of single cloud effective volumes. Using this method they extract a limiting $\tilde f_\mathrm{v,crit}\approx 0.24$ above which they find cold gas survival. This aligns with our findings for large volume filling fractions (probed by their work) but fails to predict survival for $f_v \ll 0.1$.

A handful of higher-resolution studies have turned to other setups via the introduction of more complex density fields that mimic ISM gas distributions \citep[e.g][]{bandabarragan2021,antipov2025,borodina2025}. Specifically, \citet{bandabarragan2021} perform wind-tunnel simulations with ISM initialised from a log-normal gas distribution spanning $10^4$–$10^6$ K. They examine the formation of clumpy, rain-like cold droplets for both compressive and solenoidal density fields. Although their initial conditions yield surviving cloudlets of cold gas at later times, they do not discuss the criterion for survival. Their results show a broad velocity spread for the cold phase, particularly at early times, but do not recover features such as shell formation, since the initial ISM depths remain below $\chi \rc$ (although the expected initial expansion is visible). Survival is revisited in \cite{antipov2025}, exploring a run from \cite{bandabarragan2021} and studying the survival of cold gas on an individual, one-by-one clump basis via a friends-of-friends algorithm to the cold phase domain of their box. They find on average $t_\mathrm{cool,cl} \ll t_\mathrm{cc}$, explaining the aforementioned formation of a multiphase outflow. However, their study does not explore the broader parameter space or explain the classical destruction regime where cloud-cloud interactions dominate survival and multiphase outflows emerge. Indeed, their analysis is reduced to two identical runs of $\lism \approx 50$ pc, and does not comment on the role of $f_v$ in survival.

Multiple studies have also investigated SNe–ISM interactions through vertically-stratified disk set-ups. Simulations consistently show that  the efficiency in generating 'hot' outflows depends strongly on the heighscale $h_\mathrm{SN}$ at which SNe take place \citep{martizzi2016,li2017,creasey2013}. Those exploding at high $h_\mathrm{SN}$ and therefore lower column densities drive hot outflows more efficiently. Recently \citet{vijayan2025quokka} has systematically studied the effect of SN scaleheight in the formation of a multiphase outflows. They show that the cold-mass loading factor becomes significant while the hot phase carries suffiecient specific energy to form an outflow when the equivalent heightscale for gas in the disk, $h_\mathrm{gas}$ is comparable to $h_\mathrm{SN}$. Since their simulations use $h_\mathrm{gas} \sim 1$kpc, this sweet spot can be explained from the column of ionised gas extending for a few hundreds of pc above the characteristic $h_\mathrm{gas}$. For our criterion, these set-ups with an average ISM $f_v = 0.1$ and $\lism \approx 100$ pc above $h_\mathrm{SN}$, are well above the survival threshold for multiphase outflows. However, for $h_\mathrm{SN} \gg h_\mathrm{gas}$ their work reports  outflows contain effectively no cold gas mass.
This makes intuitively sense: if there is no cold gas in a `windtunnel' setup in the first place, no multiphase outflow will emerge.
In the case of $h_{\rm SN}\gtrsim h_{\rm gas}$, one must be cautious.
 Our criterion indicates that in principle  gas layers extending only a few tens of parsecs above the supernova scale height should still form cold gas in the outflow. 
However, one must take the numerical resolution into account when interpreting these results. In order to capture the formation and growth of cold gas in the outflow, the critical cloud radius $\rcrit$ -- which can be significantly smaller than a parsec under typical ISM conditions (cf. Eq.~\eqref{eq:rcrit_values}) needs to be resolved by at least $\sim 8$ grid cells. If this condition is not met, cold gas formation will be artificially suppressed, even in cases where our criterion would otherwise predict its presence. This requirement becomes particularly important when only a small amount of cold gas is present above $h_\mathrm{SN}$, i.e., when existing clumps are intrinsically small: in such situations, under-resolving these clumps (and $\rcrit$) will prevent the condensation and growth of cold structures, leading to an underestimation of the cold mass in the simulated outflow.

Additionally, it is worth mentioning that while their study focuses exclusively on the role of $\lism$, we show that $f_v$ is also a key parameter. This distinction is particularly relevant for i.e. young stellar clusters, where SNe can create low-$f_v$ channels and subsequent explosions can have both large specific energies for the hot phase and form a significant cold gas component.

Multiple stratified disk studies include more complex physics such as self-gravity and chemical networks \citep{girichidis2016silcc,Walch2015,kim2018numerical} and report difficulties launching multiphase winds with supernovae feedback alone \citep{rathjen2025}. 
It is important to consider, that in typical ISM conditions  -- where the cold gas follow approximately a $\dd N/\dd m \propto m^{-2}$ (from $m_{\rm min,ISM}$ to $m_{\rm max}$) -- the mass fraction of clumps capable of surviving ram pressure acceleration is
\begin{equation}
    f_\mathrm{res} = \frac{\ln(m_\mathrm{max}/m_\mathrm{min,wind})}{\ln(m_\mathrm{max}/m_\mathrm{min, ISM})}
\end{equation}
where $m_{\rm min, wind}$ and $m_{\rm min, ISM}$ are the minimum clump masses that can be launched into the wind and present in the ISM, respectively.

Ideally, $m_{\rm min,wind}$ should be set by $\rcrit$. However, most large-scale simulations (e.g., stratified disk models) employ resolutions of $\Delta x\sim$ a few parsec, leaving $\rcrit$ unresolved. In this regime, $m_{\rm min, ISM}$ is determined by the cell size: $m_{\rm min,ISM}\propto \Delta x^3$. Crucially, clumps must span several cells per dimension to survive and grow via mixing in the wind. We therefore adopt $m_{\rm min,wind}\propto (8 \Delta x)^3$ to ensure adequate numerical resolution.This resolution requirement has significant consequences. Adopting an upper cutoff of $\sim 100$ pc for the maximum clump size and using $\Delta x = 4$ pc (as in \citealp{Walch2015,girichidis2016silcc}), only $f_{\rm res}\approx 0.35$ of the simulated ISM mass could be launched into the wind via ram-pressure acceleration.

For gas scale heights of $h_\mathrm{gas} \approx 100$pc \citep{Walch2015,kim2017intro}, the layers of cold gas above the supernovae scale height can extend only a fraction of $h_\mathrm{gas},\, \text{i.e.} \sim 20$pc. Combining these results, we obtain an effective cold-gas depth of $f_v(\lism f_\mathrm{res}) \approx 0.1\times0.3 \times 20\,\mathrm{pc} = 0.6$ pc $< \rcrit$ for the average density values in the ISM, which therefore does not satisfy our criterion for the formation of multiphase outflows. 
In addition, it is also worth noticing that $f_{\rm res}\lism \lesssim \Delta x$ for most stratified disk studies, and so the `effective' cold gas is not well resolved and will not launch multiphase outflows.
In other words, while supernovae detonating too deep in the disk fail to launch winds because they must propagate through too much cold gas before reaching the surface, those occurring under more favourable conditions ($\sim h_{\rm gas}$) may still not produce multiphase outflows if the amount of cold gas above the explosion site is insufficient -- or insufficiently resolved -- to seed the growth of a cold phase in the wind. Higher-resolution simulations -- which well resolve $\rcrit$ -- will shed light on this issue.

Similar arguments apply in principle to isolated disk and larger-scale simulations \citep{tanfielding,smith2021efficient}, but two additional complications arise. First, these simulations often employ adaptive mesh refinement, making it unclear whether the resolution criteria discussed above remain valid. Second, the impact of specific sink-particle and supernova-injection schemes on outflow properties remains an open question \citep{kim2018numerical}, particularly in simulations including self-gravity. Notably, high supernova rates combined with clustered star formation can efficiently drive multiphase winds even when $h_{\rm SN}\ll h_{\rm gas}$ \citep[e.g.][]{schneider2024cgols}.

Finally, although our work only partially explores the role of higher-energy winds, the AGN community has extensively studied the effect of powerful outflows in clumpy ISMs. These winds are characterised by large overdensities and wind velocities \citep{Bourne_2017,costa2014feedback}.\cite{ward2024} show that introducing clumpiness in AGN winds can alter the coupling between hot and cold phases.  In our framework, multiphase ISMs naturally lead to the formation of gas shells (see section~\ref{sec:fv_fa}).This shell structure has important implications for understanding AGN outflow driving mechanisms, as for example, a momentum-driven wind can efficiently couple to the cold gas shell through $P\dd V$ work, producing observational signatures that mimic energy-driven outflows. A complementary study by \cite{borodina2025} investigates jet propagation through multiphase ISMs and finds that the orientation and jet properties outside the disk depends on the intrinsic AGN power. At low luminosities, cold outflows are rarely produced, whereas at intermediate luminosities of order $L \sim 10^{40}\,\mathrm{erg \,s^{-1}}$, the outflow direction can be strongly altered depending on inclination and ISM depth. Although neither study reaches resolutions down to $\rcrit$, our results demonstrate that the survival criterion applies for $\chi \sim 1000$ and $v_w \sim 700$\kms  (see section~\ref{sec:overdense_examples}) and that it can account for substantial differences in the observed phase structure and energetics of AGN-driven outflows, where the energy budget is usually larger than for SNe-driven feedback.

\subsection{Caveats}
\begin{itemize}

    \item \textbf{Set-up and resolution}. Our simulations resolve individual clouds by $\gtrsim 8$ cells per radius which is sufficient to capture growth  and survival \citep{gronke2020}. Thus, computational limits restrict us to $f_v \ge 10^{-3}$. We carry out an alternative analysis indicating that survival should extend down to $f_v = 10^{-5}$ (cf. \S~\ref{app:d_clouds}), although dedicated simulations are still required to confirm this. The periodic boundary condition in the direction perpendicular to the wind is designed to mimic a larger ISM patch. Tests with varying widths (see Appendix~\ref{app:resolution}) show the results are converged.  

    \item \textbf{Magnetic fields}. Initialising magnetic fields is non-trivial and adds computational cost. Single-cloud studies \citep{mccourt2015magnetized,dursi2008draping,gronke2020} show that magnetic fields alter the picture of entrainment and morphology for single-clouds. In particular, \cite{hidalgo2024better} demonstrated that near-equipartition magnetic fields can boost survival by two orders of magnitude (for $\chi\sim 100$), shifting the effective threshold in $f_v \lism$ down to sub-parsec cloud sizes. This process -- and how it connects to a more ralistic ISM morphology -- remains incompletely understood, and further follow-up work is required.  


    \item \textbf{Outflow energetics and geometry}. We explore only part of the parameter space spanned by wind velocities, temperatures, and density contrasts in the literature. While our criterion successfully predicts cloud survival in winds with velocities up to $700 \,\mathrm{km\,s^{-1}}\, (\mathcal{M}_\mathrm{w} = 1.5)$ and contrasts of $\chi \sim 10^3$, these Mach numbers and overdensitites remain slightly below the typical conditions for SNe-ISM interactions, and well below the ones expected for AGN-ISM interactions\citep{Bourne_2017,costa2014feedback}. Extending the analysis to these regimes will be an important direction for future work. 
    As discussed in \S~\ref{sec:fv_fa}, we do not consider the change in wind properties, e.g., due to the adiabatic expansion. This will change the morphology and mass transfer rates at larger distances \citep{gronke2020,Dutta2025}. However, our core conclusions of this study are related to the initial launching, and are thus, not affected by this.

    \item \textbf{Thermal conduction and viscosity.} While single cloud studies have shown that thermal conduction and viscosity can alter the cold gas morphology and dynamics\citep{Brueggen2023}, they play and overall small effect on the mass transfer rate between the phases, and thus, the survival criterion of cold gas. This is because turbulent diffusion dominates over thermal conduction \citep{tan2021,fieldingbryan}, and rapid cooling sharpens the density (and velocity) interface, thus, counteracting the effects of viscosity \citep{marin2025limited}.
\end{itemize}

\section{Conclusions}
\label{sec:conclusions}
Galactic outflows are inherently multiphase: they regulate the cold gas content of galaxies, enrich the surrounding medium with $\sim 10^4 \, \mathrm{K}$ gas and metals, and can suppress or delay star formation. Yet, current theoretical models provide no consistent explanation for their origin, structure, and composition. In particular, the scarcity of cold gas in simulated outflows remains at odds with the observed multiphase character of galactic winds. Most attempts to address this tension have relied on single-cloud simulations, an idealised configuration that does not capture the complexity of a realistic ISM.  

In this work, we performed 3D hydrodynamic simulations of hot outflows interacting with a multicloud ISM, characterized by its cold-gas volume filling fraction $f_v$, depth $\lism$, and typical clump size $\rc$. This framework enabled us to identify the parameter regimes that naturally lead to multiphase outflows, and to assess their relevance in scale-free ISMs, which better approximate real galactic environments. Our main findings are as follows:

\begin{enumerate}
    \item \textbf{Universal multiphase outflow criterion}: Cold clumps grow for ISMs satisfying the criterion $f_v \lism \geq \rcrit$, where $\rcrit$ corresponds to the single-cloud size threshold for survival \citep{gronke2018,LiHopkins2020,Kanjilal2020}, $f_v$ is the cold gas volume filling fraction, and $\lism$ the distance the hot wind has to travel through the ISM. 
    This survival criterion can be rephrased in terms of a critical column density $N_\mathrm{crit} \gtrsim10^{18} \, \chi_\mathrm{100} \, \text{cm}^{-2}$ required for survival, analogous to the single-cloud criterion.
    This generalised survival criterion can explain the survival of cold gas clouds present, e.g., in a fractal-like ISM morphology, which would fall in the `destruction regime'  ($t_\mathrm{cool,mix} < t_\mathrm{cc}$) if considered individually. 
    We show that this criterion holds two orders of magnitude below the classical threshold, i.e., for $\rc / \rcrit \lesssim 10^{-2}$ and only breaks down at $f_v \lesssim 10^{-5}$, where intercloud distances overatake their interaction radius $d_\mathrm{int} \sim 4\chi^{1/2} \rc$.

    \item \textbf{Self-similar cold gas morphology}: Independent of the initial ISM structure, the cold phase rapidly converges toward a scale-free mass distribution following Zipf's law, $\mathrm{d}N/\mathrm{d}m \propto m^{-2}$. This behaviour emerges across all simulations and persists over time, effectively erasing memory of the initial morphology and suggesting that turbulent mixing and radiative condensation drive multiphase gas toward a universal self-similar state. As a result, the cold clump mass spectrum, rather than showing imprints of the original ISM geometry, becomes a fundamental distribution of multiphase outflows.

    \item \textbf{Compression and cold shells}: Depending on the initial conditions, the ISM depth can be compressed or expanded by the hot wind, with $\lism \lesssim \chi \rc$ leading to expansion. Surviving outflows concentrate their cold material into shells or plumes of size $d_\mathrm{cold} = \chi r_{\mathrm{cl,min}}$ along the direction of the wind, which then slowly grow as hot gas continues to be accreted.  
    
    \item \textbf{Flow and turbulence}: As the $T \sim 10^6$ K wind crosses the $T \sim 10^4$ K ISM, turbulence is driven in both phases, peaking at the ISM crossing time $t_\mathrm{sh}$. The cold and hot phases are kinematically coupled, both in bulk flow and turbulence, with the cold phase reaching $\sigma/c_\mathrm{s,cold} \approx 1$. First-order velocity structure functions show that the emergent turbulence is compatible with a Kolmogorov turbulent cascade and the injection scales are set by $\lism$.  
    
    \item \textbf{Evolving outflow properties}: The growth of multicloud outflows is governed by the Kelvin-Helmholtz instability timescale of the largest clumps, $m/\dot m \sim \chi (t_\mathrm{kh,max}t_\mathrm{cool,mix})$. Despite ISM compression, both $f_v$ and the areal covering fraction $f_A$ increase in time. $f_A$ rapidly approaches unity even for $f_v=10^{-3}$. Although $f_v$ also grows, it remains $\ll 1$, consistent observations. 
\end{enumerate}
Taken together, these results establish a framework that connects idealised single-cloud studies with stratified disk simulations, extending the analytical survival criterion to more complex, ISM morphologies.
While further work is required to test the robustness of this criterion under additional physical processes and a larger parameter space, our findings already provide a pathway for comparison with both larger-scale numerical simulations and observations of multiphase galactic outflows.

\section*{Acknowledgements}
FHP thanks Martin Fournier for help with VSF functions. FHP also thanks Tiago Costa, Aditi Vijayan and Benedikt Seidl for insightful discussions. MG thanks the Max Planck Society for support through the Max Planck Research Group, and the European Union for support through ERC-2024-STG 101165038 (ReMMU).
P.G.~acknowledges funding by the Deutsche Forschungsgemeinschaft (DFG, German Research Foundation) – 555983577.



\bibliographystyle{mnras}
\bibliography{ism_cc} 





\appendix

\section{ISM generation}
\label{app:ISM_generation}
\begin{figure}
    \centering
    \centering
    \includegraphics[width=\columnwidth]{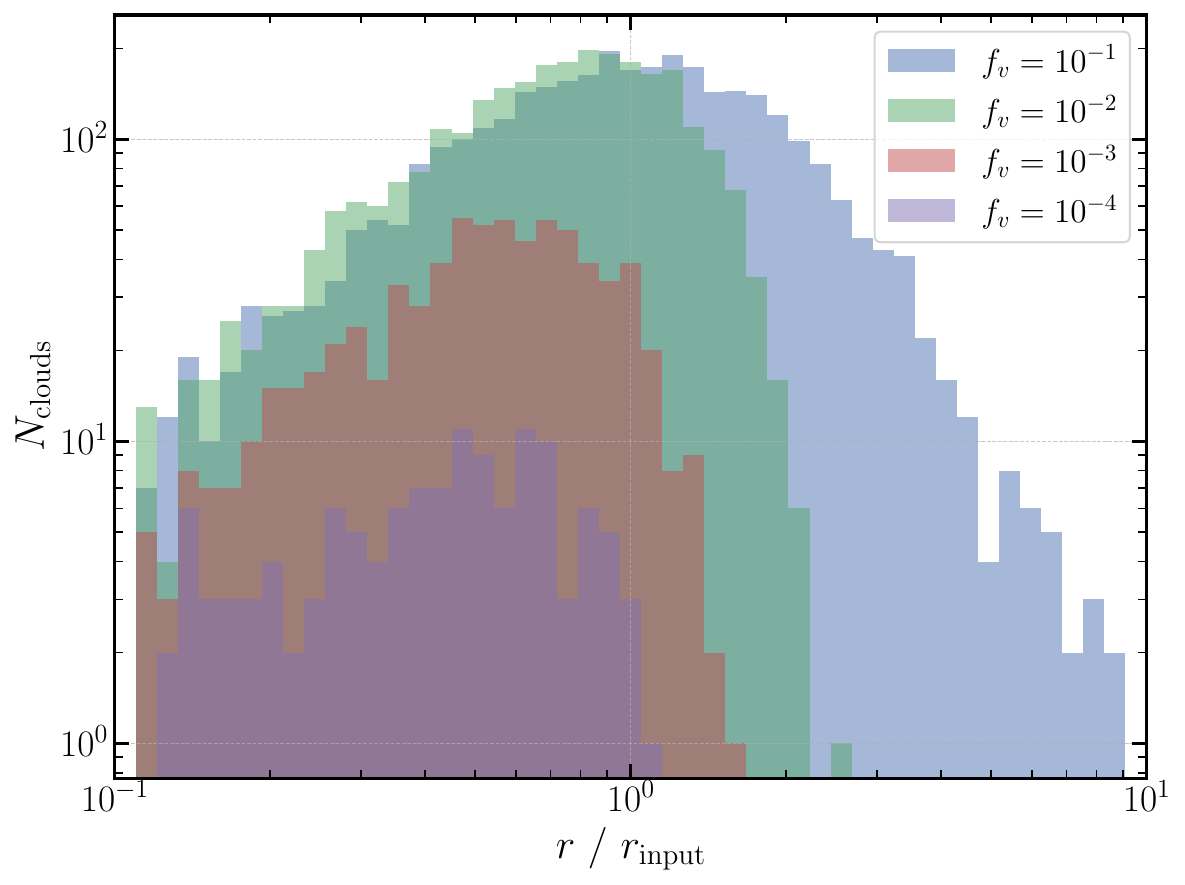}
    \caption{Clump size distribution for the ISM modelling in section~\ref{sec:methods}, shown for different volume filling fractions for a cubic box of side $10^3$ cells. The $x$ axis represents the measured clump radius as a fraction of the expected input clump size.}
    \label{fig:k_estimates}
\end{figure}

In section~\ref{sec:methods} we describe the formalism to create a binary ISM. Figure~\ref{fig:k_estimates} shows how this distribution peaks at a characteristic cloud mass scale. Lower volumetric fractions preferentially reduce the high mass clouds, leading to a slight displacement in the peak by a factor of a few. As we study variations in cloud sizes of orders of magnitude, this has negligible impact on the survival probability. This potential effect of factor of a few is in turn captured by the fudge factor of order of magnitude in eq.~\ref{eq:surv_criterion}, where $\alpha$
accounts for geometric effects.

\begin{figure}
    \centering
    \centering
    \includegraphics[width=\columnwidth]{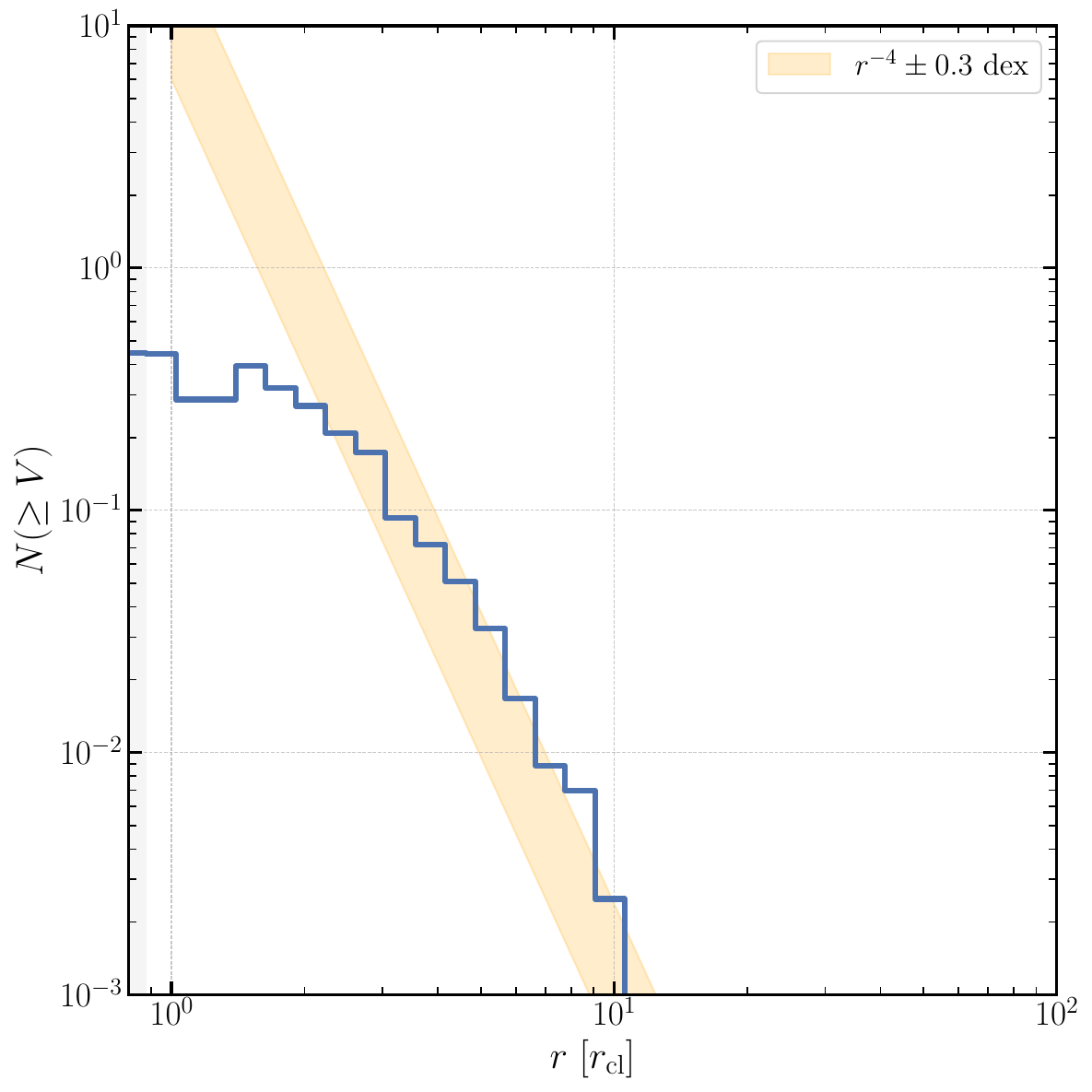}
    \caption{Cumulative distribution function of clump masses for the scale-free ISM example with overdensity $\chi = 100$ and initial parameters $(f_v, \lism/\rcrit, r_\mathrm{cl,min}/ \rcrit) = (0.1, 30  , 0.1)$. The clump size distribution spans over an order of magnitude above $r_\mathrm{cl, min}$, with a slope roughly matching Zipf's law (yellow band), where the PDF follows $\mathrm{d} N/ \mathrm{d} m \propto m^{-2}$.}
    \label{fig:scale_free_cdf}
\end{figure}

Similarly, in section~\ref{sec:methods}, we also create ISM of varying clump sizes following the mass density profile $\dd N / \dd m \propto m^{-2}$ to mimic observations of "scale-free" ISMs. We show the density profile distribution of the scale-free example in section~\ref{sec:overdense_examples} in figure~\ref{fig:scale_free_cdf} corresponding to the black-solid line in figure~\ref{Discussion/fig:overdense_evolution}. We use an overdensity of $\chi = 100$ and initial ISM parameters $(f_v, \lism/\rcrit, r_\mathrm{cl,min}/ \rcrit) = (0.1, 30  , 0.1)$. The initial clump size distribution spans over an order of above $r_\mathrm{cl, min}$, whereas the slope of the distribution roughly matches the yellow band, representing the expected Zipf's law where the PDF is proportional to $m^{-2}$.

\section{Mass convergence}
\label{app:resolution}

In order to capture the extended nature of the ISM in the direction perpendicular to the wind, we impose periodic boundary conditions for the $x$ and $z$ axis. We show that for our two fiducial resolution cases $L_\mathrm{\perp, box}/ d_\mathrm{cell} = 32$ and 16, mass growth is well converged by showing the evolution of two runs with $(\rc/\rcrit, f_v, \lism/\rc) = (1, 0.1, 30)$. Notice that we generate separate initial conditions from identical ISM parameters $(\rc/\rcrit, f_v, \lism/\rc)$. Even for this case, the mass evolution is close to identical, only showing deviations of a factor of a few around 15 $\tcc$.

\begin{figure}
    \centering
    \includegraphics[width=\columnwidth]{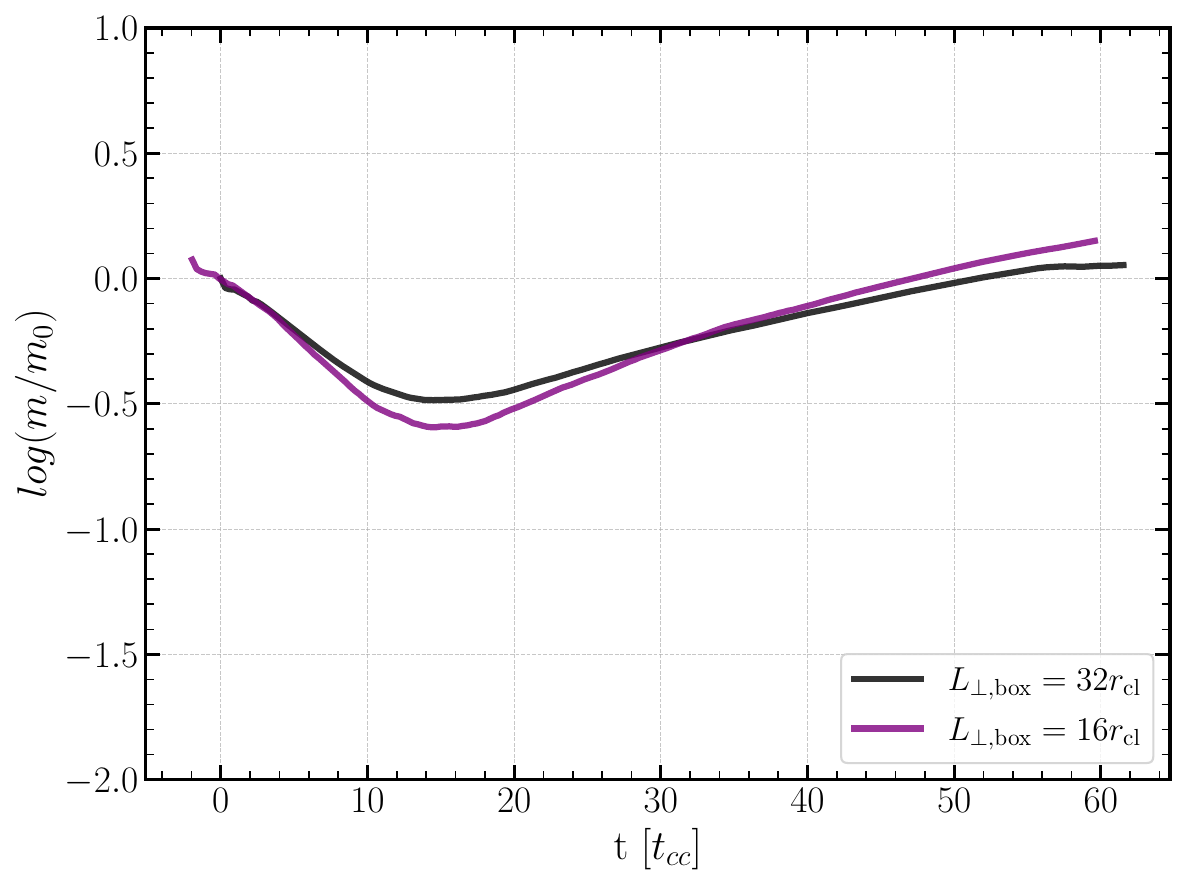}
    \caption{Mass growth for two simulations with $(\rc/\rcrit, f_v, \lism/\rc) = (1, 0.1, 30)$ for different mesh length in the direction perpendicular to the direction of propagation of the wind. In black, we use a box of $L_\perp = 128$ cells. In purple, our fiducial simulation domain with  $L_\perp = 256$. }
    \label{fig:mass_convergence}
\end{figure}

\section{Limitations of the Multiphase Survival Criterion}
\label{app:d_clouds}

\begin{figure}
    \centering
    \includegraphics[width=\columnwidth]{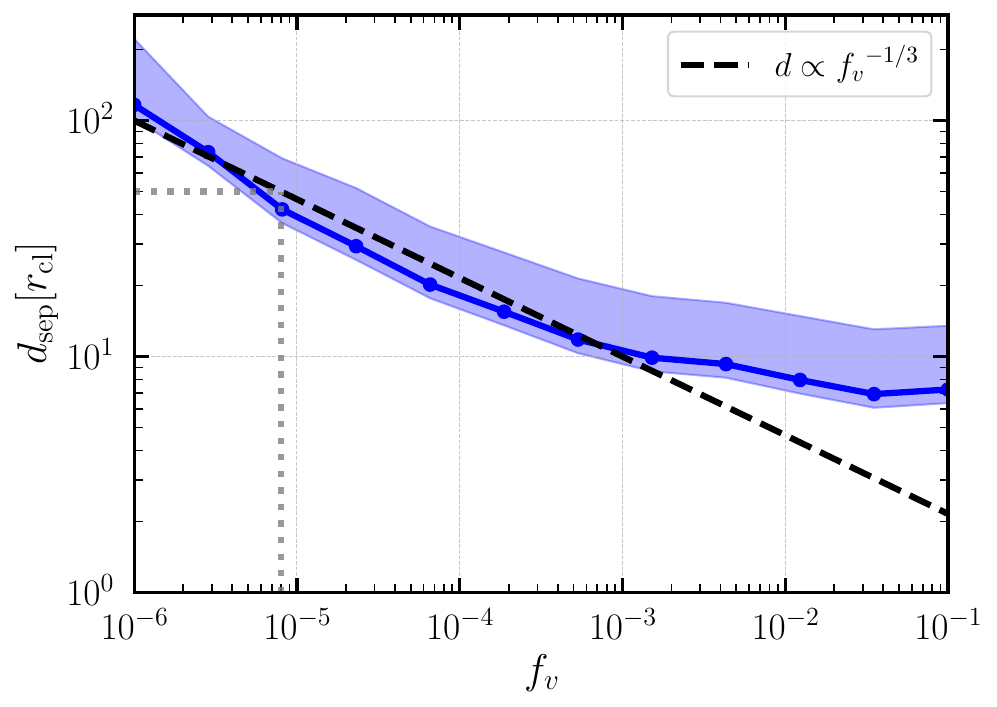}
    \caption{Average intercloud separation as a function of volume filling fraction for the cold phase. The thick dashed line shows the expected relation for a homogeneous ISM. In dotted light grey, the critical volume filling fraction for intercloud distances larger than $4\chi^{1/2}\rc$ .}
    \label{fig:cloud_separation}
\end{figure}

As demonstrated in section~\ref{sec:discussion:universal_criterion}, inter-cloud distances of $4\chi^{1/2}\rc$ mark the limit for survival in multicloud set-ups. Clouds separated by distances larger than this limit cannot interact and do not exhibit survival below $\rc < \rcrit$. We can associate this separation to a volume filling fraction of the cold phase:

Since a roughly isotropic and homogeneous ISM should have $f_v = N\rc^3/d_\mathrm{cl}$ (with $N$ as number of clouds and $d_\mathrm{cl}$ as cloud separation), the intercloud separation follows as $d_\mathrm{cl}/\rc \approx f_v^{-1/3}$, which in turn shows that the limiting radius of influence $4\chi^{1/2} \rc$, corresponds to a volume filling fraction of $\approx 10^{-5}$. The bottom panel in figure~\ref{fig:cloud_separation} shows the intercloud separation averaged for the 6 k-nearest neighbours using $125\rc^3$ ISM realisations. The isotropic cloud approximation accurately predicts $d_\mathrm{cl}$, only deviating by a factor of $\sim$ a few for volume filling fractions above $10^{-2}$. This is potentially an artifact of the k-tree and CCL algorithm (see methods of section~\ref{Results/sec:mass_distribution}), as structures overlap more at higher volume filling fractions. The prediction for the cloud separation dependence on $f_v$ agrees well  with the empirically found separation of clumps in the ISM. Indicated in dashed grey, the predicted critical separation (and therefore, volume filling fraction) for surviving clouds.

\bsp	
\label{lastpage}
\end{document}